\documentclass[sort&compress]{aipproc}

\layoutstyle{8x11single}

\usepackage[english]{babel}
\usepackage[latin1]{inputenc}

\usepackage{latexsym,amsfonts,amsmath,amsthm,amssymb}

\newcommand{\la}{\lambda}
\newcommand{\de}{\delta}

\newcommand{\Om}{\Omega}
\newcommand{\om}{\omega}
\newcommand{\F}{\cal {F}}
\newcommand{\A}{\cal {A}}
\newcommand{{\bP}}{\bf {P}}

\title{Reconstruction of quantum theory on the basis of the formula of total probability}
\author{Andrei Khrennikov}{
address={International Center for Mathematical Modeling
in Physics, Engineering and Cognitive science
MSI, V\"axj\"o University, S-35195, Sweden},
email={Andrei.Khrennikov@msi.vxu.se}
}

\begin{document}

\begin{abstract}The notion of context (complex of physical conditions) is basic in this paper.
We show that the main
structures of quantum theory (interference of probabilities, Born's rule, complex probabilistic amplitudes,  Hilbert state space, representation of observables by operators) are present in a latent form in the classical Kolmogorov probability model. However, this model should be considered as a calculus of contextual probabilities.
In our approach it is forbidden to consider abstract context independent probabilities: ``first context
and then probability.'' We start with the conventional formula of total probability for contextual
(conditional)
probabilities and then we rewrite it by eliminating combinations of incompatible contexts from consideration. In this way we obtain interference of probabilities without to appeal to the Hilbert space formalism or wave mechanics.
However, we did not just reconstruct the probabilistic formalism of conventional quantum mechanics.
Our contextual probabilistic model is essentially more general and, besides the  projection to the
complex Hilbert space,
it has other projections. The most important new prediction is the possibility (at least theoretical)
of appearance
of hyperbolic interference. A projection of the classical contextual probabilistic model to the hyperbolic
 Hilbert space
(a module over the commutative two dimensional Clifford algebra) has some similarities with the projection
 to the complex Hilbert space. However,  in the hyperbolic quantum mechanics the principle of superposition
  is violated.
Our realistic (but contextual!) approach to quantum mechanics does not contradict to various
``no-go theorems'', e.g.,
von Neumann, Bell, Kochen-Specker. We emphasize that our projection of the classical probabilistic
 model to the complex Hilbert space
is based on two incompatible observables (``reference observables''), e.g., the position and
the momentum, or the position and the energy. Only these two observables can be considered
as objective properties of quantum systems.
\end{abstract}

\maketitle

 \section{Introduction}
It is well know that the classical Kolmogorov probabilistic model [1] differs crucially
from the quantum probabilistic model, see, e.g., [2]-[37] for details and debates. The classical model
is based on a set-theoretical structure ($\sigma$-field of subsets of some set $\Omega$); there is no need in using complex numbers; physical observables are represented
 by functions  on $\Omega$ and there is no need in using noncommutative structures.
  The quantum model is based on a complex Hilbert space. The appearance of complex numbers
  in the model with real-valued probabilities and physical observables is one of quantum mysteries.
  Instead of probability distributions, there are considered complex probabilistic amplitudes
  (or in the abstract approach normalized vectors in the Hilbert state space). The probabilistic
   interpretation of such amplitudes (or vectors in the Hilbert space)  is given by {\it Born's rule.}
   It is hard to find probabilistic roots of this rule in the conventional quantum theory.
   \footnote{It is clear that M. Born wanted to modify Schr\"odinger's idea that the wave function gives
the distribution of the electric charge of electron in space. First time Born's rule appeared as a footnote
and in the first variant of the Born's paper there was  proposed to consider not square, but the absolute
value of $\psi.$} By using the Hilbert space formalism or Schr\"odinger's wave mechanics it is possible
to get {\it interference of probabilities} (which is also observed in many quantum experiments, e.g.,
the two slit experiment). There is no reasonable explanation of interference.
\footnote{Self-interference of an individual particle is a metaphor. As was pointed out by N. Bohr,
there is no way to combine the corpuscular model and interference of probabilities. This is the essence
of the principle of complementarity. But we emphasize that this principle is a consequence of one special interpretation of quantum mechanics -- the Copenhagen interpretation.} Physical observables are represented by self-adjoint operators. Operators
are noncommutative for incompatible observables. There is no explanation of the appearance of the
 noncommutative structure in the theory with real-valued (so commuting) physical observables.

These probabilistic differences between classical and quantum probabilistic models induce a rather
 mystical viewpoint
to properties of quantum systems. In this paper we shall see that the gap between
 classical and quantum probabilistic models is not so huge as it was commonly believed, see [2]-[37].
 The quantum probabilistic model can be considered as a
{\it projection of the classical model to the complex Hilbert space,} see [38]-[45]. As any projection, the quantum
projection does not provide the complete image of the prequantum classical probabilistic model.
 In particular, there can be constructed
another projection - to a so called hyperbolic Hilbert space, [38]-[45].

The notion of {\it context} (complex of physical conditions) is basic in this paper. We show that the main
structures of quantum theory (interference of probabilities, Born's rule, complex probabilistic amplitudes,
Hilbert state space, representation of observables by operators) are present in a latent form in the
classical Kolmogorov probability space.\footnote{Hence, it would be better to speak not about hidden
variables for quantum theory, but about
hidden presence of quantum structures in the classical probability space.}
However, this space should be considered as the basis of a calculus of contextual probabilities.
In our approach it is forbidden to consider abstract context independent probabilities: {\it ``first
 context and then probability.''} We start with the conventional formula of total probability for contextual (conditional)
probabilities and then we rewrite it by eliminating combinations of incompatible contexts from
consideration.\footnote{Let us imagine that in 19th (or even in 18th) century a probabilist would like to modify
the formula of total probability in such a way that ``intersections of incompatible conditions''
would be eliminated.  In such a way he
should automatically come to our formula of total probability with the interference term. Unfortunately,
 this did not happen. Interference of probabilities was discovered not in pure mathematics, but in
 experimental physics. Then it was derived by using the quantum formalism.} In this way we obtain
 interference of probabilities without to appeal to the Hilbert space formalism or wave mechanics.
By starting with the formula of total probability with interference term (under some restriction on the
magnitude of this term) we represent contexts belonging to a special class (so called trigonometric contexts) by complex probabilistic amplitudes. The Born's rule
immediately appears in such a representation. In our contextual model this rule is just a special form of
writing the
formula of total probability with the interference term.

As was already mentioned, we did not only reconstruct the probabilistic formalism of conventional quantum
 mechanics. Our contextual probabilistic model is essentially more general and, besides the projection to
  the complex Hilbert space, it has other projections. The most important new prediction is the possibility
  (at least theoretical) of appearance of {\it hyperbolic interference.} For the conventional trigonometric
  interference the brightness of interference picture is changed as a trigonometric function, e.g.,
   $\cos.$ For the hyperbolic interference the brightness of interference picture is changed as a
   hyperbolic function, e.g., $\cosh,$ so exponentially. It may occur that such an
interference with exponentially varied brightness would be found in future experiments.

Our realistic (but contextual!) approach to quantum mechanics does not contradict to various
``no-go theorems'', e.g.,
von Neumann, Kochen-Specker, Bell.  We would like to pay attention that
all business with ``no-go theorems'' is about the correspondence between two mathematical probabilistic
models: $M_{\rm{cl}}$ and $M_{\rm{quant}}.$ The main problem in the no-go activity is that nobody paid
attention to the evident fact that to study
such a mathematical problem, it is not enough  to describe two mathematical models. One should also
fix a class of rules
of the correspondence between those models. In fact, in each ``no-go theorem'' there is fixed such a class of rules.
And ``no-go'' means
only no-go for such a class of rules.  Classes of rules for classical-quantum correspondence
can be considered as classes of maps from $M_{\rm{cl}} \to M_{\rm{quant}}$ or
$M_{\rm{quant}} \to M_{\rm{cl}}.$ However, any mathematician understands well that if one proved that
there does not exist a map of some class which maps, e.g., $M_{\rm{cl}}$ into (or onto)
$M_{\rm{quant}},$ this does not mean that there could not be found a map of another
class.\footnote{Let us illustrate the situation with classical-quantum correspondence by the
following example. Let one choose the class of diffeomorphisms (i.e., one-to-one $C^1$-maps with inverse
 $C^1$-maps) as the class of correspondence-maps for configuration spaces of dynamical systems. Let he
  proved that  two configuration spaces are not diffeomorphic, i.e., it is impossible to construct
  a diffeomorphism between these spaces. But anybody understands that such a result does not mean that
  it is impossible to construct, e.g., a homeomorphism (i.e., one-to-one continuous map with inverse
   continuous) between these spaces.}

We emphasize that our projection of the classical probabilistic model to the complex Hilbert space
is based on two fixed incompatible observables ({\it ``reference observables''}), e.g., the position
 and the momentum, or the position and the energy. Only these two observables can be considered
 as {\it objective properties of quantum systems.}
In our model these observables are realized by classical incompatible  random  variables and
 incompatibility is defined in purely classical probabilistic framework, see Definition 2. Let us
 fix some pair of reference observables $a$ and $b$
taking values in the field of real numbers ${\bf R}.$ Let $C$ be some trigonometric context
(i.e., a complex of physical conditions inducing the ordinary $\cos$-interference). By using
$C$-contextual probabilities for $a$ and $b$ we represent the context $C$ by a complex probability
amplitude $\psi_C$ (this amplitude is, in fact, encoded in the formula of total probability with the
 interference term, see (\ref{TNC})). This representation induces a representation of the reference
 observables $a$ and $b$ by self-adjoint operators $\hat{a}$ and $\hat{b}.$ Incompatibility of random variables $a$ and $b$ implies that these
operators do not commute: $[\hat{a}, \hat{b}]\not=0.$

The crucial point in understanding why our contextual model does not contradict to ``no-go theorem''
of von Neumann is that the algebraic structure on the set of values of random variables, the field of real
 numbers ${\bf R},$ is not consistent with the algebraic structure on the space linear operators in the
 (complex) Hilbert state space.
For example, we shall see that in general the image of the random variable $d(\omega)=a(\omega) + b(\omega)$ is not given by the operator $\hat{d}= \hat{a}+\hat{b}.$ But the correspondence $d(\omega) \to \hat{d}$ is one of the
conditions of the von Neumann ``no-go theorem.'' This condition was criticized by many authors, see [13], [16].
 Therefore
it is not so surprising that it is violated in our model. There exist contexts $C$ such that $d(\omega)$
and  $\hat{d}$ have different probability distributions (with respect to the context $C$ and the
corresponding state
$\psi_C,$ respectively). Surprisingly, in spite of difference of probability distributions,
classical and quantum averages coincide: $E(d/C)= (\hat{d} \psi_C, \psi_C).$ The same is valid for
any random variable of the form
$d(\omega)=f(a(\omega)) + g(b(\omega))$ (and the corresponding quantum observable
$\hat{d}= f(\hat{a})+g(\hat{b})).$
Thus in our model for a wide class of ``realistic'' random variables (in particular, for any
``energy variable''
${\cal H}(\omega)= \frac{a^2(\omega)}{2 m}+ V(b(\omega))$ and the corresponding ``Hamiltonian''
$\hat{{\cal H}}= \frac{\hat{a^2}}{2 m}+ V(\hat{b}))$ quantum averages coincides with prequantum
classical averages (so, in particular, $E({\cal H}/C)= (\hat{{\cal H}}\psi_C, \psi_C)).$

The existence of our realistic prequantum model does not contradict to ``no-go theorem'' of Kochen-Specker,
since our model does not define  a one-to-one map from the space of quantum observables into the space
of classical random variables. In this paper we do not consider composite systems. Therefore we do not
discuss relations with
Bell's theorem.

\section{Interference of  probabilities}

\subsection{The conventional Kolmogorov probabilistic model}

Let ${\cal K}=(\Om, \F, {\bP})$ be a Kolmogorov probability space, [1], [46]. This space is the  basis
of the classical probabilistic model, the Kolmogorov model [1], [46]. As any model of reality, the Kolmogorov model
consists of two parts: the mathematical formalism and the interpretation.

{\it Mathematical formalism.} Here $\Omega$ is an arbitrary set, ${\cal F}$ is a
$\sigma$-field\footnote{A collection of subsets of $\Omega$ which contains $\Omega$ and the empty
set $\emptyset$ and it is closed with respect to the operations of  countable intersection and union
of sets and it contains the complement to any its element.} of subsets of $\Omega;$ ${\bf P}$ is
a probability measure on ${\cal F}:$ a countably-additive measure with values in $[0,1]$ such
that ${\bf P}(\Omega)=1.$

{\it Kolmogorov's interpretation.} Points  $\omega\in \Omega$ represent {\it elementary events}.
Some special sets of
elementary events represent {\it events}; it is supposed that the family of all events   is
a $\sigma$-field ${\cal F}.$ For an event $A\in {\cal F},$  ${\bf P}(A)$ is the probability of
{\it occurrence of the event} $A.$ Observables (e.g., physical) are represented by {\it random variables.}
 We recall that a random variable is a  measurable function $d: \Omega \to {\bf R}$ (so for any Borel
 subset $\Gamma$ of the field of real numbers ${\bf R},$ its preimage
 $D_\Gamma= \{ \omega \in \Omega: d(\omega) \in \Gamma\}$ belongs to the $\sigma$-field
${\cal F}).$ Conditional probability ${\bf P}(B/A)$ that an event $B$ occurs under the condition that
an event $A$
has been occurred is defined by {\it Bayes' formula:}
\begin{equation}\label{BAY}
{\bf P}(A/C)=\frac{{\bf P}(A\cap C)}{{\bf P}(C)}, {\bf P}(C) \ne 0.
\end{equation}

\subsection{The contextual Kolmogorov probabilistic model}

Here we use the same mathematical formalism as in the conventional Kolmogorov probabilistic model, the
Kolmogorov probability space ${\cal K}=(\Om, \F, {\bP}).$ However, structures of ${\cal K}$ have
different interpretations.

{\it Contextual interpretation of the Kolmogorov probability space.} Points $\omega \in \Omega$
represent {\it fundamental parameters} of the model.\footnote{We recall that we would not like to call
$\omega$ hidden variables, since we are not looking for hidden parameters for the quantum model.
We are looking for the hidden quantum structure in the Kolmogorov probability space ${\cal K}.$}
Some special sets of fundamental parameters represent {\it contexts} -- complexes
of physical conditions.\footnote{In this paper we consider only physical models. However,
 it is possible to use the same
approach for, e.g., cognitive or psychological models, see [41], [47].} It is supposed that sets
representing contexts
form a $\sigma$-field ${\cal F}.$ In the opposite to the conventional Kolmogorov probabilistic model,
${\bf P}(C), C \in {\cal F},$  has no direct physical interpretation. In our model probability can be considered only as
conditional (or better to say contextual) probability, see (\ref{BAY1}). As in the conventional
 Kolmogorov probabilistic model, observables are represented  by random variables. For a random
 variable $d,$ the conditional (contextual) probability ${\bf P}(d \in \Gamma/C), C \in {\cal F},$ is
 defined by the Bayes' formula:
\begin{equation}\label{BAY1}
{\bf P}(d \in \Gamma/C) =\frac{{\bf P}(D_\Gamma \cap  C)}{{\bf P}(C)}, {\bf P}(C) \ne 0.
\end{equation}
In our model the Bayes' formula has the following meaning.
To find the probability that a random variable $d\in \Gamma$ under the context $C,$ there should be selected parameters $\omega \in \Omega$
which belong the intersection of the sets $D_\Gamma$ and $C.$ These are all parameters $\omega \in C$ for that
$d(\omega) \in \Gamma.$
We remark that the Bayes' formula (\ref{BAY1}) gives the definition of probability in terms of
the contextual Kolmogorov model.  The ``experimental probability'' ${\bf P}_{\rm{exp}}(d \in \Gamma/C)$
is defined as the limit of the frequencies $\nu_N(d \in \Gamma/C)$  to find $d \in \Gamma$ in a series
of $N$ observations under the complex of physical conditions $C.$\footnote{According to [48] there
two levels of description of physical reality:
ontic and epistemic. Kolmogorov probabilities ${\bf P}(d \in \Gamma/C)$ belong to the ontic
level and frequency probabilities ${\bf P}_{\rm{exp}}(d \in \Gamma/C)$ belong to the epistemic level.}

   The ``experimental probability'' ${\bf P}_{\rm{exp}}(d \in \Gamma/C)$
coincides with ${\bf P}(d \in \Gamma/C)$ as a consequence of the {\it law of large numbers}
(if trails are independent).
We have  the same situation in the conventional Kolmogorov model.

In particular, if $d$ is a discrete random variable then
\begin{equation}\label{BAY2}
{\bf P}(d =z \in \Gamma/C) =\frac{{\bf P}(D_z \cap  C)}{{\bf P}(C)}, {\bf P}(C) \ne 0,
\end{equation}
where $D_z=\{\omega \in \Omega: d(\omega)=z\}.$

By our interpretation the set $D_z$ {\it represents
the context corresponding to filtration with respect to value} $d=z.$ We emphasize that that the
operation of intersection of sets has nothing to do with with creating ``intersections'' of corresponding
contexts. The probability
${\bf P}(D_z \cap  C)$ has no physical interpretation by itself.\footnote{We remark that any model
of physical reality contains some mathematical structures which do not have direct physical interpretations.
 For example, A. N. Kolmogorov
pointed out that the condition of countable-additivity do not permit physical verification, [1].
There can also exist measurable sets which do not correspond to physical events and so on.}
 We shall discuss this point in more detail in the next section.

\subsection{The formula of total probability}

As was remarked in introduction, our visualization of the latent quantum structure of the classical
contextual
probabilistic model is based on a contextual version of the well known formula of total probability.
 We start
with recalling this formula in the conventional Kolmogorov approach.

Let ${\A}=\{ A_n \}$ be finite or countable {\it complete group of disjoint events} (``partition of unity''):
$$
A_i \cap A_j= \emptyset, i \not= j,\;\;\;\;  \cup_i A_i=\Om.$$
Let $B, C\in {\cal F}$  be events and let ${\bP}(C)>0.$
We have the standard formula of total probability, see, e.g., [46]:
\begin{equation}
\label{CONV}
{\bP}(B/C)=\sum_n {\bP}(A_n/C){\bP}(B/A_n \cap C)
\end{equation}
which can be easily derived:
$$
{\bP}(B/C)=\frac{{\bP}(B\cap C)}{{\bP}(C)}=
\sum_n \frac{{\bP}(B\cap A_n\cap C) {\bP}(A_n \cap C)}{{\bP}(C) {\bP}(A_n \cap C)}.
$$
This derivation was performed under the condition that
\begin{equation}
\label{DQM}
{\bP}(A_n \cap C)>0 \; \mbox{for all}\; n.
\end{equation}
A contextual analog of this condition will play an important role in our theory.
The formula (\ref{CONV}) works well in all domains of science (it is the basis of Bayesian analysis),
besides quantum physics (and may be psychology, see [41], [47]).
In particular, let $a$ and $b$ be discrete random variables taking values
$a \in Y=\{a_i, i=1, \ldots , k_a\}$ and $b \in X=\{b_j, j=1, \ldots, k_b\},$ where $k_a, k_b < \infty.$ We have
\begin{equation}
\label{CONV1}
{\bP}(b=x /C)=\sum_{y \in Y} {\bP}(a=y/C) {\bP}(b=x/ (a=y)\cap C), x \in X \;.
\end{equation}
For further considerations it is useful to introduce sets:
$$
A_y =\{ \omega \in \Omega: a(\omega)=y\}, y \in Y, \;\;  \; B_x=\{\omega \in \Omega: b(\omega)=x\}, x\in X.
$$

\subsection{The formula of total probability with interference term}

We now want to consider this formula in the contextual Kolmogorov model. Since the mathematical formalism
is the same,
there are no differences in mathematical calculations; the only difference is in the interpretation.
Probabilities  ${\bP}(b=x /C), {\bP}(a=y/C)$ are well defined from the contextual viewpoint. Let us now
consider the probability \\ ${\bP}(b=x/ (a=y)\cap C)= {\bP}(b=x/ A_y \cap C).$ Here considerations are
not so straightforward. In the conventional Kolmogorov model the set  $Q=(a=y) \cap C = A_y \cap C$
represents the event -- the simultaneous occurrence of the events $A_y$ and $C.$
To perform careful analysis of the contextual situation, at the moment we shall use different symbols
for a context
and the corresponding set in ${\cal F}$ representing this context: contexts will be denoted
$\tilde{C}, \tilde{A_y},
\tilde{Q},...$ and corresponding sets $C, A_y,Q,...$ In particular, here $\tilde{A_y}$ is the context
of the $[a=y]$-filtration which is represented by the set $A_y$ in the Kolmogorov space.

In our model the set $Q=A_y \cap C$ represents some context $\tilde{Q}(y,C).$ But the representation of the
 set $Q$ in the form of the intersection of the sets $C$ and $A_y$ does not mean that the context
  $\tilde{Q}(C,y)$ is really the ``intersection'' of the contexts $\tilde{C}$ and
   $\tilde{A_y}.$\footnote{There is no such a postulate in our interpretation of the probability space
   ${\cal K}.$}  The latter context, say $\tilde{M}(y,C)$ --
{\it ``first we prepare an ensemble of systems under the complex of physical conditions $C$ and then
perform the $a=y$ filtration''} -- need not be represented by the set $Q.$ If the procedure of  $a=y$
filtration disturbs the original context $C,$ then there is no reason to assume that the context
$\tilde{M}(y,C)$ should be represented by the set $Q=A_y \cap C.$ So the first conclusion of our analysis
is that in general the sets $Q=A_y\cap C$ in the right-hand side of
(\ref{CONV1}) do not represent contexts $\tilde{M}(y,C).$ We remark that the contexts $\tilde{M}(y,C)$
can be easily designed experimentally and used for the collection of statistical data for frequency probabilities
(which can be found in a long series of observations). What can we say about a context $\tilde{Q}$ which
is represented by the set $Q?$ In fact, not so much. This context should be created via the
nondisturbative $[a=y]$-selection under the complex of physical conditions $C.$ In general $\tilde{Q}$
cannot be constructed just through the combination of $\tilde{C}$ and $\tilde{A_y}.$

Since we do not know how to create the context $\tilde{Q},$  we would not be able to find corresponding
experimental probabilities and the formula of total probability is not useful for applications (in spite
of its validity
in the underlying Kolmogorov model). \footnote{By using terminology of [48] one can say that the formula
of total probability is well defined on the ontic level of description of nature, but it could not be
directly lifted to the epistemic level of description. We would like to modify this formula to get its
 analog which would be meaningful
on both levels of description.} Therefore it would be natural to try to exclude sets $Q=A_y\cap C$ from
consideration and obtain a new variant of the formula of total probability.\footnote{We emphasize that
 we do not claim
that the context $\tilde{Q}$ represented by $Q=A_y\cap C$ could not be created at all, cf. with Bohr's
principle of complementarity and some interpretations of Heisenberg's  uncertainty relations. We
only observed that the conventional formula of total probability (\ref{CONV1}) contains sets
$Q=A_y\cap C$ representing contexts $\tilde{Q}$ such that
in general we do not know how to create them. Therefore we would like to exclude sets $Q=A_y\cap C$ from
our consideration. Finally, we shall come to the same formalism that is used in quantum mechanics. But we
shall escape
a lot of quantum mysteries, since sets $Q=A_y\cap C$ are eliminated from formulas by a simple pragmatic
 reason.}
Our analysis of correspondence between  creating new contexts and operations on sets representing
contexts is finished.
We shall again use the same symbol for a context and the set representing this context.

To simplify considerations, we shall consider only {\it dichotomous random variables}:
$a\in Y=\{a_1, a_2,\},b \in X=\{b_1,b_2\}.$ Even this very simple model (the contextual Kolmogorov model with dichotomous observables) has (in a latent form) all distinguishing features of the quantum model.

{\bf Definition 1.} (cf. (\ref{DQM}) {\it{A context $C \in \F$ is  nondegenerate  with respect to a
random variable $a$ if ${\bP}(A_y \cap C)\not =0$ for all $y\in Y.$}}

We denote the set of all $a$-nondegenerate contexts by the symbol  ${\cal C}_{a}.$

{\bf Definition 2.} {\it{ Random variables $a$ and $b$ are called incompatible if
${\bP}(A_y \cap B_x ) \not = 0$ for all $y \in Y$ and $x\in X.$}}

Thus $a$ and $b$ are incompatible iff every $B_x$ is $a$-nondegerate and vice versa. We introduced
incompatible
random variables in purely classical framework (see appendix for some properties of such pairs
of random variables). We shall see that this incompatibility contains
(in a latent form)
quantum incompatibility -- representation by noncommutative operators.

{\bf Theorem 1.} (Formula of total probability with interference term)
 {\it Let $a$ and $b$ be incompatible random variables and let a context $C$ be $a$-nondegenerate.
 Then }
\begin{equation}
\label{INN}
{\bP}(b=x/C)=\sum_{y\in Y} {\bP}(a=y/C){\bP}(b=x/a=y)+
2 \lambda(b=x/a,C)\sqrt{\prod_{y\in Y}{\bP}(a=y/C)
{\bP}(b=x/a=y)},
\end{equation}
where
\begin{equation}
\label{TNCT2}
\lambda(b=x/a, C)=
\frac{\delta(b=x/a,C)}{2\sqrt{{\bP}(a=a_1/C) {\bP}(b=x/a=a_1){\bP}(a=a_2/C) {\bP}(b=x/a=a_2)}}
\end{equation}
and
\begin{equation}
\label{D1}
\delta(b=x/a,C) ={\bf P}(b=x/C) - \sum_{y\in Y} {\bf P}(b=x/a=y){\bf P}(a=y/C).
\end{equation}

To prove Theorem we put expressions for $\lambda$ and $\delta$ into
(\ref{INN}) and we obtain identity. In fact, (\ref{INN}) is just a representation of
the probability ${\bP}(b=x/C)$ in a special way. We choose the special representation of the
perturbation $\delta(b=x/a, C)$ of $\sum_{y\in Y} {\bP}(a=y/C){\bP}(b=x/a=y),$
namely its normalization by square root of all probabilities. At the beginning I expected that
this normalization would produce quantities bounded by one. But in general it was not the case.
We call $\lambda(b=x/a, C)$ {\it coefficients of incompatibility.}
In our further investigations we will use the following result:

{\bf Lemma 1.} {\it{Let conditions of Theorem 1 hold true. Then}}
\begin{equation}
\label{CD1}
\sum_{x\in X} \de(b=x/a,C)=0
\end{equation}

{\bf Proof.} We have $1=\sum_{x\in X} {\bP}(b=x/C)=\sum_{x\in X} \sum_{y\in Y} {\bP}(a=y/C)
{\bP}(b=x/a=y)+\sum_{x\in X} \de(b=x/a,C).$
But $\sum_{y\in Y}(\sum_{x\in X} {\bP}(b=x/a=y)) {\bP}(a=y/C)=1.$

As a consequence of this lemma we have:
\begin{equation}
\label{CD2}
\sum_{x\in X} \la(b=x/a,C)\sqrt{{\bP}(a=a_1/C){\bP}(a=a_2/C)
{\bP}(b=x/a=a_1) {\bP}(b=x/a=a_2)}=0 .
\end{equation}

\medskip

1). Suppose that  both coefficients  of incompatibility are relatively small
\[\vert \la(b=x/a,C)\vert\leq 1, \; x \in X.\]
In this case we can introduce new statistical parameters $\theta(b=x/a,C)\in
[0,2 \pi]$ and represent the coefficients of statistical disturbance in the
trigonometric form:
\begin{equation}
\label{TNCT}\la(b=x/a,C)=\cos \theta (b=x/a,C).
\end{equation}
Parameters $\theta(b=x/a,C)$ are said to be {\it{relative phases}} (or random variables $a$ and $b.)$
This is purely probabilistic definition of phases. So we
introduce geometry through probability.

In this case we obtain the following interference formula of total probability:
\begin{equation}
\label{TNC}{\bP}(b=x/C)=\sum_{y\in Y} {\bP}(a=y/C){\bP}(b=x/a=y)+
2 \cos(b=x/a,C)\sqrt{\prod_{y\in Y}{\bP}(a=y/C){\bP}(b=x/a=y)} .
\end{equation}
This is nothing other than the famous {\it formula of interference of
probabilities.}\footnote{Typically this formula is derived by using the Hilbert
space (unitary) transformation corresponding to the transition from
one orthonormal basis to another and Born's probability postulate.
The orthonormal basis under quantum consideration consist of eigenvectors of
operators (noncommutative) corresponding to quantum physical observables
$a$ and $b.$} Thus we found (hidden) interference of probabilities in the
Kolmogorov probability space.

2). Suppose that  both coefficients  of incompatibility are relatively large
\[\vert \la(b=x/a,C)\vert\geq 1, \; x \in X.\]
In this case we can introduce new statistical parameters $\theta(b=x/a,C))\in
(-\infty ,+ \infty)$ and represent the coefficients of incompatibility in the
hyperbolic form:
\begin{equation}
\label{TNCT1}
\la(b=x/a,C)=\pm \cosh \theta(b=x/a,C).
\end{equation}
Parameters $\theta(b=x/a,C)$ are said to be hyperbolic {\it{relative phases}}.
In this case we obtain the following interference formula of total probability:
\begin{equation}
\label{TNC1}
{\bP}(b=x/C)=\sum_{y\in Y} {\bP}(a=y/C){\bP}(b=x/a=y)\pm
2 \cosh(b=x/a,C)\sqrt{\prod_{y\in Y}{\bP}(a=y/C)
{\bP}(b=x/a=y)}
\end{equation}
We remark that in the ordinary formula for interference of probabilities (\ref{TNC}) the expression in
 the right-hand side determines the quantity  which belongs the segment [0,1] for any angle $\theta.$ In
 the hyperbolic case, see
(\ref{TNC1}), this quantity belongs  [0,1] only for special range of angles $\theta.$ But this is not a
problem
in our approach. We do not determine the probability ${\bP}(b=x/C)$ through the right-hand side of
(\ref{TNC1}).
We proceed in the opposite way: the phase $\theta$ is determined through probabilities ${\bP}(b=x/C),
{\bP}(a=y/C), {\bP}(b=x/a=y).$ For some probabilities there exists the  trigonometric representation,
 for other probabilities there exists the hyperbolic representation.

3). Suppose that the absolute value of one of the coefficients $\la(b=x/a,C)$ is less than one
and the  absolute value of another coefficient is larger than one.
Here we have the interference formula of total probability containing
trigonometric as well as hyperbolic interference terms.

If incompatible random variables $a$ and $b$ are fixed,
we shall often use the symbols  $\delta(x/a, C), \lambda(x/a, C)$
instead of $\delta(b=x/a, C), \lambda(b=x/a, C).$

\section{Quantum projection of the classical model}

Let us fix a pair of incompatible random variables $a=a_1, a_2, b=b_1, b_2.$
We call such variables {\it reference variables.} For each pair $a, b$ of reference variables,
we construct a projection of the contextual Kolmogorov model to the complex Hilbert space.
We start from the {\it trigonometric interference.}
We set
$$
{\cal C}^{\rm{tr}} =\{ C\in {\cal C}_{a}: \vert\lambda(b=x/a, C)\vert\leq 1\}
$$
We call elements of ${\cal C}^{\rm{tr}}$ trigonometric contexts. We shall see that quantum mechanics can be interpreted
as a representation of trigonometric contexts. We  shall also consider hyperbolic contexts
which can be represented in a hyperbolic Hilbert space.
In  few further sections  we shall consider only
trigonometric contexts and in those sections we shall omit the upper index and use simply the symbol:
${\cal C}\equiv{\cal C}^{\rm{tr}}.$

\subsection{Interference and complex probability amplitude, Born's rule}

Let $C\in {\cal C}.$  We set
$p_C^a(y)={\bP}(a=y/C), p_C^b(x)={\bP}(b=x/C), p(x/y)={\bP}(b=x/a=y),
x \in X, y \in Y.$
The interference formula of total probability (\ref{TNC}) can be written in
the following form
\begin{equation}
\label{Two}
p_C^b(x)=\sum_{y \in Y}p_C^a(y) p(x/y) + 2\cos \theta_C(x)\sqrt{\Pi_{y \in
Y}p_C^a(y) p(x/y)}\;,
\end{equation}
where $\theta_C(x)=\theta(b= x/a, C)= \pm \arccos \lambda(b=x/a, C), x \in X, C\in {\cal C}.$ Here
$$
\delta(b=x/a, C)=p_c^b(x)-\sum_{y \in Y} p_C^a(y)p(x/y)\;\; \mbox{and}\;\;
\lambda(b=x/a, C)  =\frac{\delta(b=x/a, C)}{2\sqrt{\Pi_{y\in Y}p_C^a(y)p(x/y)}} .
$$
By using the elementary formula:
$$
D=A+B+2\sqrt{AB}\cos \theta=\vert \sqrt{A}+e^{i \theta}\sqrt{B}|^2,
$$
for $A, B > 0, \theta\in [0,2 \pi],$
we can represent the probability $p_C^b(x)$ as the square of the complex amplitude (Born's rule):
\begin{equation}
\label{Born}
p_C^b(x)=\vert\psi_C(x)\vert^2 \;,
\end{equation}
where
\begin{equation}
\label{EX1}
\psi(x)\equiv \psi_C(x)=\sqrt{p_C^a(a_1)p(x/a_1)} + e^{i \theta_C(x)} \sqrt{p_C^a(a_2)p(x/a_2)} \;.
\end{equation}

It is important to underline that since  for each $x\in X$ phases $\theta_C(x)$ can be
chosen in two ways (by choosing signs + or -) a representation of contexts by complex
amplitudes is not uniquely defined.\footnote{To fix a representation of a contextual
Kolmogorov space ${\cal K}$ we should fix phases. We shall see that to obtain a
``good representation'' we should choose phases in a special way.}

We denote the space of functions: $\psi: X\to {\bf C}$ by the symbol
$E=\Phi(X, {\bf C}).$ Since $X= \{b_1, b_2 \},$ the $E$ is the two dimensional
complex linear space. Dirac's $\delta-$functions $\{ \delta(b_1-x), \delta(b_2-x)\}$
form the canonical basis in this space. We shall see
that under a natural restriction
on the matrices of transition probabilities:
$$
\psi_{B_z}(x)= \delta(z - x), z=b_1, b_2.
$$
For each $\psi \in E$ we have $\psi(x)=\psi(b_1) \delta(b_1-x) + \psi(b_2) \delta(b_2-x).$
By using the representation (\ref{EX1}) we construct the map
\begin{equation}
\label{MAP}
J^{b/a}:{\cal C} \to \Phi(X, {\bf C})
\end{equation}
The $J^{b/a}$ maps contexts (complexes of, e.g., physical conditions) into complex
amplitudes. The representation ({\ref{Born}}) of probability as the square of the
absolute value of the complex $(b/a)-$amplitude is nothing other than the
famous {\it Born rule.}

{\bf Remark.} {\small We underline that the complex linear space representation (\ref{EX1}) of the
set of contexts ${\cal C}$ is based on a pair $(a,b)$ of incompatible
(Kolmogorovian) random variables. Here $\psi_C=\psi_C^{b/a}.$ We call random variables
$a, b$ {\it reference variables.} Such a pair of variables determines a ``probabilistic
system of coordinates'' on a contextual Kolmogorov space.}

The complex amplitude $\psi_C(x)$ can be called a {\it wave function} of the
complex of physical conditions, context $C$  or a  (pure) quantum {\it state.}
In principle, we can represent each context $C \in {\cal C}$ by a family of complex amplitudes:
\begin{equation}
\label{EX}\psi(x)\equiv \psi_C(x)=\sum_{y\in Y}\sqrt{p_C^a(y)p(x/y)}
e^{i \xi_C(x/y)}
\end{equation}
such that $\xi_C(x/a_1) - \xi_C (x/a_2)=\theta_C(x) .$
For such complex amplitudes we also have Born's rule (\ref{Born}).
However, to simplify considerations we shall consider only the representation (\ref{EX1}) and the map
(\ref{MAP}) induced by this representation.

\subsection{Hilbert space representation of  the $b$-variable}

We set $e_x^b(\cdot)=\delta(x- \cdot).$ For any context $C\in {\cal C},$ the complex amplitude
$\psi_C$ can be expanded as:
\begin{equation}
\label{EIV}
\psi_C= \sum_{x\in X} \psi_C(x) e_x^b
\end{equation}
Thus the Born's rule for complex amplitudes (\ref{Born}) can be rewritten in the following form:
\begin{equation}
\label{BH}
p_C^b(x)=\vert(\psi_C, e_x^b)\vert^2 \;,
\end{equation}
where the scalar product in the space $E=\Phi(X, C)$ is defined by the
standard formula:
\begin{equation}
\label{BHS}
(\psi, \psi) = \sum_{x\in X} \psi(x)\bar \psi(x).
\end{equation}
The system of functions $\{e_x^b\}_{x\in X}$ is an orthonormal basis in the
Hilbert space $H=(E, (\cdot, \cdot)).$

Let $X \subset {\bf R}.$ By using the Hilbert space representation ({\ref{BH}}) of
the Born's rule  we obtain  the Hilbert space representation of the
expectation of the (Kolmogorovian) random variable $b$:
\begin{equation}
\label{BI1}
E (b/C)= \sum_{x\in X}xp_C^b(x)=\sum_{x\in X}x\vert\psi_C(x)\vert^2=
\sum_{x\in X}x (\psi_C, e_x^b) \overline{(\psi_C, e_x^b)}=
(\hat b \psi_C, \psi_C) \;,
\end{equation}
where  the  (self-adjoint) operator $\hat b: H \to H$ is determined by its
eigenvectors: $\hat b e_x^b=x e^b_x, x\in X.$
This is the  multiplication operator in the space of complex functions $\Phi(X,{\bf C}):$
$$
\hat{b} \psi(x) = x \psi(x)
$$
 By (\ref{BI1}) the  conditional expectation of the Kolmogorovian
random variable $b$ is represented
with the aid of the self-adjoint operator $\hat b.$  Therefore it is natural to
represent this random variable (in the Hilbert space model)  by the operator $\hat b.$
So the Hilbert space image $\hat{b}$ of the random variable $b$ was defined through the formula
(\ref{BI1}) for conditional average. This formula is a simple consequence of the Born's rule
(\ref{BH}). And the Born's rule is present in a latent form in the formula of total probability
with interference term (\ref{TNC}). This formula induces the representation of a context $C$ by the
complex amplitude $\psi_C$ defined by  (\ref{EX1}). The amplitude has  a natural expansion with respect
to the basis $\{e^b_x\}_{x\in X},$ see (\ref{EIV}). And this basis induces the representation (\ref{BI1}).

We would like to introduce an operator $\hat{a}$ representing the random variable $a$ by using similar arguments.
But we emphasize that random variables $a$ and $b$ do not play the same role in the Hilbert space
representation under consideration. In fact, we now consider the $b/a$-projection of ${\cal K}.$

\subsection{Born's rule for the $a$-variable}

We start with the complex amplitude $\psi_C$ defined by (\ref{EX1}). We shall see that this amplitude
can be expanded  with respect to a natural basis, $\{e^a_y\}_{y\in Y}.$ That expansion
plays the role similar to the expansion (\ref{EIV}) with respect to the
basis $\{e^b_x\}_{x\in X},$ namely  Born's rule takes place
for the $a$-variable (under a natural restriction to the matrix of transition probabilities):
\begin{equation}
\label{BBR}
p_C^a(y)=\vert(\psi, e_y^a)\vert^2, \; y \in Y .
\end{equation}
We set:
\begin{equation}
\label{KOE}
u_j^a=\sqrt{p_C^a(a_j)}, u_j^b=\sqrt{p_C^b(b_j)}, p_{ij}=p(b_j/a_i), u_{ij}=\sqrt{p_{ij}},
\theta_j=\theta_C(b_j), e^b_j= e^b_{b_j},  e^a_j= e^a_{a_j}.
\end{equation}
We remark that the coefficients $u_j^a, u_j^b$ depend on a context $C;$ so
$u_j^a=u_j^a(C), u_j^b=u_j^b(C).$
We also consider the {\it matrix of transition
probabilities} ${\bf P}^{b/a}=(p_{ij}).$
It is always a {\it stochastic matrix.}\footnote{So $p_{i1}+p_{i2}=1, i=1,2.$}
We have, see (\ref{EIV}), that
$$
\psi_C=v_1^b e_1^b + v_2^b e_2^b, \;\mbox{where}\;\;
v_j^b=u_1^a u_{1j}  + u_2^a u_{2j} e^{i \theta_j}\;.
$$
Hence
\begin{equation}
\label{BI}
p_C^b(b_j) =\vert v_j^b \vert^2 = \vert u_1^a u_{1j}  + u_2^a u_{2j} e^{i \theta_j} \vert^2.
\end{equation}
This is the {\it interference representation of probabilities} that is used,
e.g., in quantum formalism.\footnote{ By starting with the general representation (\ref{EX})
we obtain $v_j^b=u_1^a u_{1j} e^{i\xi_{1j}} + u_2^a u_{2j} e^{i\xi_{2j}}$ and the interference
representation $p_C^b(b_j)=\vert v_j^b\vert^2=\vert u_1^a u_{1j} e^{i\xi_{1j}} + u_2^a
u_{2j} e^{i\xi_{2j}}\vert^2.$}

For any context $C_0,$ we can represent the corresponding wave function
$\psi=\psi_{C_0}$ in the form:
\begin{equation}
\label{0}
\psi=u_1^a e_1^a + u_2^a e_2^a,
\end{equation}
where
\begin{equation}
\label{Bas}
e_1^a= (u_{11}, \; \; u_{12}) ,\; \;
e_2^a= (e^{i \theta_1} u_{21}, \; \; e^{i \theta_2} u_{22})
\end{equation}
We suppose that vectors $\{e_i^a\}$ are linearly independent, so $\{e_i^a\}$
is a basis in $H.$ We have:
\[e_1^a=v_{11} e_1^b + v_{12} e_2^b, \; \; \; e_2^a=v_{21} e_1^b + v_{22} e_2^b \]
Here $V=(v_{ij})$
is the matrix:
$v_{11}=u_{11}, v_{21}=u_{21}$ and $v_{12}=e^{i \theta_1} u_{21}, v_{22}=
e^{i \theta_2} u_{22}.$
We would like to find a class of matrixes $V$ such that
Born's rule (\ref{BBR}) holds.
By (\ref{0}) we have the
Born's rule (\ref{BBR}) iff $\{e_i^a\}$ was an {\it orthonormal
basis,} i.e., the  $V$ was a {\it unitary} matrix. Since we study the
two-dimensional case (i.e., dichotomous random variables), $V\equiv
V^{b/a}$ is unitary iff the matrix of transition probabilities ${\bf
P}^{b/a}$ is {\it double stochastic} and $e^{i\theta_1}=-e^{i\theta_2}$ or
\begin{equation}
\label{MAR}
\theta_{C_0}(b_1) - \theta_{C_0}(b_2)=\pi \mod 2\pi
\end{equation}
We recall that a matrix is double stochastic if it is stochastic, i.e.,
$p_{j1} + p_{j2}=1,$ and, moreover,
\begin{equation}
\label{DSC}
p_{1j} + p_{2j}=1, j=1,2.
\end{equation}
Double stochasticity is equivalent to the condition:
$p_{11}=p_{22}, p_{12}=p_{21}.$ Any matrix of transition probabilities is stochastic
(as a consequence of additivity of the conditional probability),
but in general it is not double stochastic.
We remark that the constraint (\ref{MAR}) on phases and the double stochasticity constraint
(\ref{DSC}) are not independent:

{\bf Lemma 2.} {\it Let $a$ and $b$ be incompatible random variables and let the
matrix of transition probabilities ${\bf P}^{b/a}$ be double stochastic.
Then:
\begin{equation}
\label{SW}
\cos \theta_C (b_2)=-\cos \theta_C (b_1)
\end{equation}
for any context $C\in {\cal C}.$}

{\bf Proof.} By Lemma 1 we have:
\[\sum_{x\in X}\cos\theta_C(x)\sqrt{\Pi_{y\in Y} p_C^a(y) p(x/y)}=0\]

But for a double stochastic matrix ${\bf P}^{b/a}=(p(x/y))$ we have:
\[\Pi_{y\in Y} p_C^a (a_1) p(b_1/y)=
\Pi_{y\in Y} p_C^a (a_2) p(b_2/y) .\]
Since random variables $a$ and $b$ are incompatible, we have $p(x/y)\not =
0, x\in X, y\in Y.$ Since $C\in {\cal C}_{a},$ we have $p_C^a(y)\not =
0, y\in Y.$ We obtain (\ref{SW}).

By Lemma 2 we have two different possibilities to choose phases:
\[\theta_{C_0}(b_1) + \theta_{C_0}(b_2) = \pi  \;\rm{or} \;
\theta_{C_0}(b_1) - \theta_{C_0}(b_2) = \pi \mod 2\pi\]
By (\ref{MAR}) to obtain the Born's rule for the $a$-variable we should choose phases $\theta_{C_0}(b_i), i=1,2,$ in such a way that
\begin{equation}
 \label{MAR0}
 \theta_{C_0}(b_2)=\theta_{C_0}(b_1) + \pi.
\end{equation}
If $\theta_{C_0}(b_1)\in [0, \pi]$ then $\theta_{C_0}(b_2)\in [\pi, 2\pi]$ and vice versa.
Lemma 2 is very important since by it (in the case when reference observables are chosen
in such way that the matrix of transition probabilities is double stochastic)
we can always choose $\theta_{C_0}(b_j), j=1,2,$ to satisfy (\ref{MAR0}).

The delicate feature of the presented construction of the $a$-representation is that the basis $e_y^a$ depends on the context $C_0: e_y^a=e_y^a(C_0).$ And the Born's rule, in fact, has the form:
$$p_{C_0}^a(y)=|(\psi_{C_0}, e_y^a(C_0))|^2.$$
We would like to use (as in the conventional quantum formalism)
one fixed $a$-basis for all contexts $C\in {\cal C}.$ We may try to use for all contexts
$C\in {\cal C}$ the basis $e_y^a\equiv e_y^a(C_0)$ corresponding to one fixed context $C_0.$ We shall
see that this is  really the  fruitful  strategy.

{\bf Lemma 3.}
{\it {Let ${\bf P}^{b/a}$ be double
stochastic and let for any context $C\in {\cal C}$ phases $\theta_C(b_j)$ be chosen as
\begin{equation}
\label{P}
\theta_C(b_2)=\theta_C(b_1) + \pi \mod 2\pi.
\end{equation}
Then for any context $C\in {\cal C}$ we have the Born's rule (\ref{BBR}) for the basis $e_y^a\equiv e_y^a(C_0)$
constructed for a fixed context}} $C_0.$

{\bf Proof.} Let $C_0\in {\cal C}.$ We take the basic
$\{ e_y^a(C_0)\}$ (and the matric $V(C_0))$ corresponding to this context.
For any $C \in {\cal C},$ we would like to represent the wave function $\psi_C$ as
\begin{equation}
\label{LUU}
\psi_C= v_1^a(C) e_1^a(C_0) + v_2^a(C) e_2^a(C_0),\;\; \mbox{where}\;\;\; \vert v_j^a(C) \vert^2= p_C^a(a_j).
\end{equation}
It is clear that, for any $C \in {\cal C},$ we can represent the wave function as
$$
\psi_C(b_1) = u_1^a(C) v_{11}(C_0) + e^{i[\theta_C(b_1)- \theta_{C_0}(b_1)]} u_2^a(C) v_{12}(C_0)
$$
$$
\psi_C(b_2) = u_1^a(C) v_{21}(C_0) + e^{i [\theta_C(b_2)- \theta_{C_0}(b_2)]} u_2^a(C) v_{22}(C_0)
$$
Thus to obtain (\ref{LUU}) we should have:
\begin{equation}
\label{LUU1}
\theta_C(b_1)- \theta_{C_0}(b_1)=
\theta_C(b_2)- \theta_{C_0}(b_2)   \mod  2\pi
\end{equation}
for any pair of contexts $C_0$ and $C_1.$
By using the relations (\ref{P}) between phases
$\theta_C(b_1), \theta_C(b_2)$ and $\theta_{C_0}(b_1), \theta_{C_0}(b_2)$ we obtain:
$$
\theta_C(b_2)-\theta_{C_0}(b_2)=(\theta_C(b_1) + \pi-\theta_{C_0}(b_1)-\pi)=
\theta_C(b_1)-\theta_{C_0}(b_1) \mod 2\pi.
$$

\medskip

The constraint (\ref{P}) essentially restricted the class of complex amplitudes which can
be used to represent a context $C\in {\cal C}$. Any $C$ can be represented only by
two amplitudes $\psi(x)$ and $\bar{\psi}(x)$ corresponding to the two possible
choices of $\theta_C(b_1):$ in $[0, \pi]$ or $(\pi, 2\pi$).

By Lemma 3 we obtain the following result playing the fundamental role in our approach:

{\bf Theorem 2.} {\it We can construct the complex Hilbert space representation of
the contextual Kolmogorov probability model such that the Born's rule holds for both reference
variables iff the matrix of transition probabilities ${\bf P}^{b/a}$ is double stochastic.}

If ${\bf P}^{b/a}$ is double stochastic we have the quantum representation not only for
the classical conditional expectation of the variable $b,$ see (\ref{BI1}), but also for the variable $a:$
\begin{equation}
\label{EXP}
E(a/C) = \sum_{y\in Y} y p_C^a(y)= \sum_{y\in Y} y \vert(\psi_C, e_y^a)\vert^2
=(\hat {a}\psi_C, \psi_C) \;,
\end{equation}
where the self-adjoint operator (symmetric matrix)
$\hat{a} :H \to H$ is determined by its eigenvectors: $\hat{a}
e_y^a= y e_y^a.$  By (\ref{EXP}) it is natural to represent the random variable
$a$ by the operator $\hat{a}.$ Of course, the representation of random variables by linear operators is
just a convenient mathematical tool to represent the average of a random
variable by using only the Hilbert space structure.

Let us denote the unit sphere in the Hilbert space $H$
by the symbol $S.$ The map $J^{b/a}:{\cal C}\to S$ need not be a surjection (injection).
In general the set of (pure) states corresponding to a contextual Kolmogorov
space
$
S_{\cal  C}\equiv S^{b/a}_{\cal  C}
=J^{b/a}({\cal C})
$
is just a proper subset of the sphere $S.$ The structure of the set of pure states
$S_{\cal C}$ is determined by the Kolmogorov space.

\subsection{Some properties of the quantum projection}

Let ${\bf P}^{b/a}$ be double stochastic and let phases be chosen according to (\ref{P}).

The contexts $A_y$ are degenerate with respect to the $a$-variable,
since ${\bf P}(A_{a_1} \cap A_{a_2})=0$.
Thus $J^{b/a}(A_i)$ cannot be defined by (\ref{EX1}).
It is natural to extend the map  $J^{b/a}$ to sets $A_y$
by setting
$
J^{b/a}(A_y)=e_y^a, y \in Y.
$
We set
$$
\overline{{\cal C}}={\cal C} \cup {\cal A}, \; {\cal A}= \{A_{a_1}, A_{a_2}\}.
$$
Thus we have constructed the Hilbert space representation:
$J^{b/a}:\overline{\cal C}\to S.$
We set $S_{\overline{\cal C}}=J^{b/a}\overline{\cal C}.$

Let $\delta(x/a, C)=0, i=1,2.$ \footnote{We remark that
by Lemma 1 the sum of perturbation coefficients $\delta(x/a, C)$ is always
equal to zero. Thus those coefficients are equal to zero or distinct from zero at the same
time.} Here $\lambda(x/a, C)=0$ and hence (for $x \in X):$
$\theta_C(b_1)=\frac{\pi}{2}$ or $\theta_C(b_1)=\frac{3}{2}\pi.$ In the first case we have
$$
\psi_C(b_1)=\sqrt{p_C^a(a_1) p(b_1/a_1)} + i\sqrt{p_C^a(a_2)p(b_1/a_2)}
$$
\begin{equation}
\label{MAR3}
\psi_C(b_2)=\sqrt{p_C^a(a_1) p(b_2/a_1)} - i\sqrt{p_C^a (a_2)p(b_2/a_2)}
\end{equation}
The second choice of phases gives the representation of $C$ by the complex amplitude
which is conjugate to (\ref{MAR3}). We set
$$
{\cal C}_0= \{ C \in {\cal C}: \delta(x/a, C)=0, x \in X\}.
$$
We remark that $\Omega$ always belong to ${\cal C}_0.$
However, in general ${\cal C}_0\not= \{ \Omega\}.$
By considering contexts $C\in{\cal C}_0$ we would not find any sign of the latent quantum structure
in the classical probability space. But it should be underlined that
${\cal C}_0\equiv {\cal C}_0(a,b).$ Thus there can exist another pair of incompatible rrandom variables,
$a^\prime, b^\prime$ such that they produce nontrivial interference for a context $C \in {\cal C}_0(a,b).$

Let $C_1, C_2 \in {\cal C}$ be contexts such that probability distributions
of random variables $a$ and $b$ under $C_1$ and $C_2,$ respectively, coincide:
$$
p_{C_1}^a(y)= p_{C_2}^a(y), y \in Y, \; \;
p_{C_1}^b(x)=p_{C_2}^b(x), x \in X.
$$
In such a case $\lambda(x/ a, C_1) = \lambda(x/a, C_2)$ and
$\theta(x/ a, C_1)= \pm \theta(x/ a, C_2).$
If there is such a coincidence of probability distributions
for only a pair of contexts $(C_1, C_2),$ then we can
represent $C_1$ and $C_2$ by two different complex
amplitudes, $\psi_{C_2}=\bar \psi_{C_1}.$ But if we have the
coincidence for a triple of contexts $(C_1, C_2, C_3)$ then it would be
impossible to represent them by different complex amplitudes. We should
choose $\psi_{C_3}=\psi_{C_1}$ or
$\psi_{C_3}=\psi_{C_2};$ so $J^{b/a}(C_3)=J^{b/a}(C_1)$ or $J^{b/a}(C_3)=J^{b/a}(C_2).$
Thus in general the map $J^{b/a}$ is  not injective.

\section{Nonquantum Hilbert space projections of the contextual Kolmogorov
model}

Of course, for arbitrary random variables $a$ and $b$ the matrix ${\bf P}^{b/a}$ need
not be double stochastic. In this case we could not obtain Born's rule both for the $b$ and
$a$ variables. In general, for each random variable we should introduce its own scalar
product and corresponding Hilbert space: $H_b=(E, (\cdot, \cdot)_b), H_a=(E, (\cdot, \cdot)_a), \ldots, $ where
$
(\psi, \phi)_b=\sum_{j}v_j^b \bar w_j^b\;\mbox{for}\; \;
\psi=\sum_j v_j^b e_j^b, \phi=\sum_j w_j e_j^b,
$
and
$
(\psi, \phi)_a =\sum_j v_j^a \bar w_j^a \; \mbox{for} \; \psi=\sum_j v_j^a e_j^a, \phi=\sum_j w_j^a e_j^a.
$
The Hilbert spaces
$H_b, H_a,...$ give the $b-$representation, the $a-$representation, $\ldots.$
Thus $p_C^b(x)=\vert(\psi, e_x^b)_b\vert^2$ and $p_C^a(y)
=\vert(\psi, e_y^a)_a\vert^2$ and so on.

However, the cruicial difficulty is that, as we have already discussed,
$e_y^a=e_y^a(C_0)$ and, in fact, for any context $C_0\in {\cal C}$ we constructed
its own Hilbert space representation for the $a$-variable: $H_a=H_a(C_0).$
In the same way as in the above considerations
we would be able to use the same representation
for contexts $C$ and $C_0$ if the condition (\ref{LUU1}) holds true. Thus we should have:
$$
\theta_C(b_2)=\theta_C(b_1)+\alpha  \; \rm{and}\;
\theta_{C_0}(b_2)=\theta_{C_0}(b_1)+\alpha \mod 2\pi ,
$$
where $\alpha$ is some phase (if ${\bf P}^{b/a}$ is double stochastic then $\alpha=\pi$).

{\bf Theorem 3.}{\it{ Suppose that ${\bf P}^{b/a}$ is not
double stochastic and ${\cal C}\ne{\cal C}_0.$} Then there is no such an $\alpha$ that
\begin{equation}
\label{MARV}
\theta_C(b_2)=\theta_C(b_1)+\alpha
\end{equation}
for all contexts} $C\in{\cal C}.$

To prove this theorem we need the following generalization of Lemma 2:

{\bf Lemma 2a.} {\it Let $a$ and $b$ be incompatible random variables. Then
for any context $C\in {\cal C}$ the following equality holds true:
\begin{equation}
\label{T}
\cos \theta_{C}(b_2)=-k \cos \theta_C (b_1)
\end{equation}
where}
\[k\equiv k^{b/a}=\sqrt{\frac{p_{11}p_{21}}{p_{12}p_{22}}}\]

It is also easy to obtain:

\medskip

{\bf Proposition 1.} {\it The coefficient $k^{b/a}=1$ iff ${\bf P}^{b/a}$ is double
stochastic.}

\medskip

{\bf Proof of Theorem.}
By Lemma 2a we have: $-k \cos \theta_C(b_1)=\cos(\theta_C(b_1)+\alpha)$ We take $C=\Omega$ and obtain:
$\cos(\theta_\Omega(b_1)+\alpha)=0.$
But $\theta_\Omega(b_1)=\pm \frac{\pi}{2}.$ Thus $\theta_\Omega(b_1)+\alpha=\pm\frac{\pi}{2}$
and $\alpha=0, \pi \mod 2\pi.$

Since ${\cal C}\ne {\cal C}_0$ there exists a context $C$ such that $\cos \theta_C(b_1)\ne 0.$
If $\alpha=0$ then $\cos \theta_C(b_1)(k+1)=0.$ This contradicts to the positivity of $k.$
Let $\alpha=\pi.$ Then $\cos \theta_C(b_1)(k-1)=0.$ Thus $k=1.$ But this implies (by Proposition 1)
that ${\bf P}^{b/a}$ is double stochastic.

\medskip

Despite Theorem 3, we can still hope that there can be found
some extended family ${\cal C}^\prime$ of contexts such that (\ref{MARV})
would hold true for contexts $C\in {\cal C}^\prime.$ But it is impossible:

{\bf Proposition 2.} {\it{ Let condition (\ref{MARV}) hold true for two contexts $C_1, C_2$ such that
\begin{equation}
\label{NM}
|\lambda(b_1/a, C_1)|\ne|\lambda(b_2/a, C_2)|.
\end{equation}
Then ${\bf P}^{b/a}$ is double stochastic.}}

{\bf Proof.}
We set $\theta=\theta_{C_1}(b_1)$ and $\theta^\prime=\theta_{C_2}(b_1).$
We have: $-k\cos \theta=\cos (\theta+\alpha), -k \cos \theta^\prime=\cos (\theta^\prime + \alpha).$ Thus
$$
\small -k\cos\frac{\theta + \theta^\prime}{2} \cos\frac{\theta-\theta^\prime}{2}
=\cos\left(\frac{\theta + \theta^\prime}{2} + \alpha\right) \cos\frac{\theta-\theta^\prime}{2}.
$$
By (\ref{NM}) we have that $\cos \frac{\theta-\theta^\prime}{2}\ne 0$
and hence $-k \cos \frac{\theta + \theta^\prime}{2}=\cos (\frac{\theta + \theta^\prime}{2}+\alpha).$

We also have
$$
k\sin \frac{\theta + \theta^\prime}{2} \sin \frac{\theta-\theta^\prime}{2}=
-\sin\left(\frac{\theta + \theta^\prime}{2}+\alpha\right) \sin\frac{\theta-\theta^\prime}{2}.
$$
By (\ref{NM}) we have that $\sin\frac{\theta-\theta^\prime}{2}\ne 0$ and hence
$-k \sin \frac{\theta + \theta^\prime}{2}=\sin(\frac{\theta + \theta^\prime}{2} + \alpha).$
Thus $k^2=1$ and hence $k=1.$ By proposition 1 the matrix ${\bf P}^{b/a}$ is double stochastic.

Thus if ${\bf P}^{b/a}$ is not double stochastic then every surface
$M_t= \{C \in {\cal C}:|\lambda(b_1/a, C)|= t\}, 0 \leq t \leq 1,$
in the space of contexts is represented in its own Hilbert space $H_a(t).$

\section{Noncommutativity of operators representing Kolmogorovian random variables}

Let
${\bf P}^{b/a}$ be double stochastic and let phases be chosen according to (\ref{P}).
We consider in this section the case of real valued random variables. Here
spectra  of random variables $b$ and $a$ are subsets of ${\bf R}.$
We set $q_1= \sqrt{p_{11}}=\sqrt{p_{22}}$ and $q_2= \sqrt{p_{12}}= \sqrt{p_{21}}.$
Thus the vectors of the $a$-basis, see (\ref{Bas}), have the following form:
$$
e_1^a= (q_1, q_2), \; \; e_2^a= (e^{i \theta_1} q_2, e^{i \theta_2} q_1)\;.
$$
Since $\theta_2 = \theta_1+ \pi,$ we get $e_2^a= e^{i \theta_2} (- q_2,  q_1).$
We now find matrices of operators $\hat{a}$ and $\hat{b}$ in the $b$-representation. The latter one
is diagonal. For $\hat{a}$ we have:
$\hat{a}= V \rm{diag}(a_1, a_2) V^\star,$ where $v_{11}=v_{22}=q_1, v_{21}=-v_{12}=q_2.$ Thus
$
a_{11}= a_1q_1^2 +a_2 q_2^2, \; a_{22}= a_1q_2^2 +a_2 q_1^2,\;
a_{12} = a_{21}= (a_1-a_2) q_1 q_2 .
$
Hence
$$
[\hat{b}, \hat{a}] = \hat{m},
$$
where $m_{11}=m_{22}=0$ and $m_{12}=- m_{21}= (a_1- a_2) (b_2-b_1) q_1 q_2.$
Since $a_1\not= a_2, b_1\not= b_2$ and $q_j\not=0,$ we have $\hat{m}\not=0.$

\section{The role of simultaneous double stochasticity of ${\bf P}^{b/a}$ and
${\bf P}^{a/b}$}

Starting with the $b$-representation -- complex amplitudes $\psi_C(x)$ defined on
the spectrum (range of values) of a random variable $b$ -- we
constructed the $a$-representation. This construction is natural (i.e., it  produces the
Born's probability rule) only if the  ${\bf P}^{b/a}$ is double stochastic.
We would like to have a symmetric model. So by starting with
the $a$-representation -- complex amplitudes $\psi_C(y)$ defined on
the spectrum (range of values) of a random variable $a$ -- we would like
to construct the natural $b$-representation. Thus both matrices of transition
probabilities ${\bf P}^{b/a}$ and ${\bf P}^{a/b}$ should be double stochastic.

We set $B_j= B_{b_j}, A_j= A_{a_j}, j=1,2.$ It is assumed that phases are always chosen according to (\ref{P}).

{\bf Theorem 4.} {\it Let the matrix ${\bf P}^{b/a}$ be double stochastic. The contexts
$B_1, B_2$ belong to ${\cal C}$ iff the matrix ${\bf P}^{a/b}$ is double stochastic.}

{\bf Proof.} We have
$$
\lambda(B_2/a, B_1) = -\frac{\mu_1^2 +\mu_2^2}{2\mu_1 \mu_2},
$$
where $\mu_j= \sqrt{p^a_{B_1}(a_j) p(b_2/a_j)}.$ So $\lambda(B_2/a, B_1)\geq 1$
and we have the trigonometric behavior only in the case $\mu_1= \mu_2.$ Thus:
$
p^a_{B_1}(a_1) p(b_2/a_1)=p^a_{B_1}(a_2) p(b_2/a_2).
$
In this case $\lambda(B_2/a, B_1)= -1,$ so we can choose, e.g. $\theta(B_2/a, B_1)= \pi,$
and consequently $\theta(B_1/a, B_1)=0.$ We pay attention to the fact that
$p^a_{B_i}(a_j)= p^{a/b}(a_j/b_i)\equiv p(a_j/b_i).$ Thus we have:
\begin{equation}
\label{EQW}
p(a_1/b_1) p(b_2/a_1)= p(a_2/b_1) p(b_2/a_2).
\end{equation}
In the same way by using conditioning with respect to $B_2$ we obtain:
$
p(a_1/b_2) p(b_1/a_1)= p(a_2/b_2) p(b_1/a_2).
$
By using double stochasticity of ${\bf P}^{b/a}$ we can rewrite the last
equality as
\begin{equation}
\label{EQW1}
p(a_1/b_2) p(b_2/a_2)= p(a_2/b_2) p(b_2/a_1).
\end{equation}
Thus by (\ref{EQW}) and (\ref{EQW1}) we have:
$$
\frac{p(a_1/b_2)}{p(a_2/b_1)} =\frac{p(a_2/b_2)}{p(a_1/b_1)}.
$$
Hence $p(a_1/b_2) = t p(a_2/b_1)$ and $p(a_2/b_2)= t p(a_1/b_1), t >0.$
But $1= p(a_1/b_2) +p(a_2/b_2)= t[ p(a_2/b_1) + p(a_1/b_1)] =t.$

To finish the proof, we need the following well known result:

{\bf Lemma 4.} {\it Both matrices of transition probabilities ${\bf P}^{b/a}$ and ${\bf P}^{a/b}$
are double stochastic iff the transition probabilities are symmetric, i.e.,
\begin{equation}
\label{SYM}
p(b_i/a_j)=p(a_j/b_i), i, j=1,2 .
\end{equation}
This is equivalent that random variables $a$ and $b$ have the uniform probability distribution:
$$p^a(a_i)=p^b(b_i)=1/2, i=1,2.$$}

This Lemma has important physical consequences. A natural (Bornian) Hilbert space representation
of contexts can be constructed only on the basis of a pair of (incompatible) uniformly distributed
random variables.

{\bf Lemma 5.}
{\it{Let both matrices ${\bf P}^{b/a}$ and ${\bf P}^{a/b}$ be double stochastic. Then}}
\begin{equation}
\label{L}
\la (B_i/a, B_i) =1 .
\end{equation}

\noindent
{\bf Proof.} Here $\delta(B_i/a, B_i)=1-p(b_i/a_1) p(a_1/b_i)-p(b_i/a_2) p(a_2/b_i)
=1-p(a_1/b_i)^2- p(a_2/b_i)^2 = 2p(a_1/b_i) p(a_2/b_i).$ Thus
$
\lambda(B_i/a, B_i)=1.
$
\bigskip
By (\ref{L}) we have \[\la (B_i/a, B_j)=-1, i \not = j,\]
Thus
\[\theta (B_i/a, B_i)=0\; \mbox{and} \;\theta (B_i/a, B_j)=\pi, i \not = j.\]

\noindent

{\bf Proposition 3.}
{\it{Let  ${\bf P}^{b/a}$ and ${\bf P}^{a/b}$ be double stochastic. Then}}
$$
J^{b/a}(B_j)(x)=\delta(b_j-x), x  \in X, \;\; \mbox{and} \; \; \; J^{a/b}(A_j)(y)=\delta(a_j-y), y \in Y.
$$
\noindent
{\bf Proof.}
Because $\theta(B_1/a, B_1)=0$ we have:
$$
J^{b/a}(B_1)(b_1)=
\sqrt{p(a_1/b_1) p(b_1/a_1)} + e^{i0} \sqrt{p(a_2/b_1) p(b_1/a_2)}
$$
$$
=
p(a_1/b_1)+ p(a_2/b_1)=1.
$$
Because $\theta (B_2/a, B_1)=\pi$ we have
$$
J^{b/a}(B_1)(b_2)=\sqrt{p(a_1/b_1) p(b_2/a_1)}
+ e^{i\pi} \sqrt{p(a_2/b_1) p(b_2/a_2)}
$$
$$
=
\sqrt{p(a_1/b_1)}(\sqrt{p(b_2/a_1}-\sqrt{p(a_2/b_1)})=0.
$$
Thus in this case:
$J^{b/a}(B_i)=e^b_i ,  i=1,2.$

Thus in the case when both matrices of transition probabilities
${\bf P}^{a/b}$ and ${\bf P}^{b/a}$ are double stochastic (i.e., both reference
variables $a$ and $b$ are uniformly distributed) the Born's rule has the form:
\begin{equation}
\label{BM}
p_C^b(x) = \vert (\psi_C, \psi_{B_x})\vert^2, \; \; p_C^a(y) = \vert (\psi_C, \psi_{A_y})\vert^2.
\end{equation}

In principle, we could start directly with constructing a quantum-like representation
of the constextual Kolmogorovian model in the case of uniformly distributed reference
variables. In this case the whole construction looks more natural. But we started
with the general representation based on an arbitrary pair of incompatible reference
variables to see how general the formalism can be, cf. Mackey [14].

\section{Example}

We consider an example of a Kolmogorov probability space and a pair of dichotomous random
variables $a, b$ which are incompatible. In this example
the set of contexts with nontrivial disturbance term $\delta, \delta \not = 0,$ is nonempty,
so ${\cal C}_0 \not= {\cal C}.$ We find the image $S_{\cal C}$ of the set of contexts $\cal C$ in the Hilbert sphere
$S \subset H.$ In this example $S_{\cal C}$ is a proper subset
of the sphere $S.$ The Hilbert space representation map $J^{b/a}$ is not injective.
Random variables $a$ and $b$ are represented by symmetric
operators in the Hilbert space $H.$ They do not commute.

Let $\Omega=\{\omega_1, \omega_2, \omega_3, \omega_4\}$ and ${\bf P}(\om_j)=p_j>0, \sum_{j=1}^4 p_j=1.$ Let
\[A_1=\{\omega_1, \omega_2\}, A_2=\{\omega_3, \omega_4\}\]
\[B_1=\{\omega_1, \omega_4\}, B_2=\{\omega_2, \omega_3\}\]

Let $p_1=p_3=q < \frac{1}{2}$ and $p_2=p_4=(1-2q)/2.$ We denote this Kolmogorov
probability space by the symbol ${\cal K}(q).$

Here ${\bf P}(A_1)={\bf P}(A_2)={\bf P}(B_1)={\bf P}(B_2)=\frac{1}{2}.$ So the random
variables $a$ and $b$ are uniformly distributed. Thus both matrices of transition probabilities
${\bf P}^{b/a}$ and ${\bf P}^{a/b}$ are double stochastic. Here
\[{\bf P}^{b/a}={\bf P}^{a/b}= \left( \begin{array}{lr}
2q  & 1-2q\\
1-2q & 2q
\end{array}
\right) \]
We have the symmetry condition ${\bf P}(B_i/A_j)={\bf P}(A_j/B_i).$

We start with two-points contexts.

(a) Let $C=C_{24}=\{\omega_2, \omega_4\}.$ Here ${\bf P}(C)=1-2q, {\bf P}(B_j/C)={\bf P}(A_j/C)=\frac{1}{2}.$
Thus $\delta=0.$ By using the representation (\ref{MAR3}) and choosing
$\theta_{C_{24}}(b_1)=\frac{\pi}{2}, \theta_{C_{24}}(b_2)=\frac{3\pi}{2}$ we obtain:
\begin{equation}
\label{T_1}
\psi_{C_{24}}(x)= \left\{ \begin{array}{ll}
{{\sqrt{q}} + i \sqrt{\frac{1-2q}{2}}, x=b_1}\\
{\sqrt{\frac{1-2q}{2}} - i\sqrt{q}, x=b_2}
\end{array}
\right .
\end{equation}

(b). Let $C=C_{13}=\{\omega_1, \omega_3\}.$ Here everything is as in (a).
We set $\theta_{C_{13}}(b_1)=\frac{3}{2} \pi$ and $\theta_{C_{13}}(b_2)=\frac{\pi}{2}.$ Thus
\[ \psi_{C_{13}}(x)=
 \left \{ \begin{array}{cc}
\sqrt{q}-i \sqrt{\frac{1-2q}{2}}, x=b_1\\
\sqrt{\frac{1-2q}{2}} + i \sqrt{q}, x=b_2
\end{array}
\right.
\]
We remark that
 $\psi_{C_{24}} \perp \psi_{C_{13}}:$
\[(\psi_{C_{24}}, \psi_{C_{13}})=
\left(\sqrt{q} + i\sqrt{\frac{1-2q}{2}}\right)^2 + \left(\sqrt{\frac{1-2q}{2}}-i\sqrt{q}\right)^2=0\]

(c) Let $C=C_{14}=\{\omega_1, \omega_4\}=B_1.$ By general theory we have
$\psi_{C_{14}}(x)=\delta(b_1-x)=e_1^b.$
In the same way: $\psi_{C_{23}}=\delta(b_2-x)=e_2^b.$

To find the Hilbert space representation of sets $C=C_{12}=\{\omega_1, \omega_2\}=A_1$ and
$C=C_{34}=\{\omega_3, \omega_4\}=A_2$ we have to construct the basis $\{e_j^a\}.$
We can choose:
\[e_1^a= \left( \begin{array}{cc}
{\sqrt{2q}}\\
{\sqrt{1-2q}}
\end{array}
\right)
\; \; e_2^a= i \left( \begin{array}{cc}
{-\sqrt{1-2q}}\\
{\sqrt{2q}}
\end{array}
\right)\]
Here we have chosen $\theta_2= \theta_{C_{13}}(b_2) = \frac{\pi}{2}.$ Thus we shall
use the $a$-basis corresponding to the context $C_0\equiv C_{13}.$ We have $\psi_{C_{12}}=e_1^a,
\psi_{C_{34}}=e_2^a.$  In a) and b) we found the probabilistic amplitudes representing contexts
$C_{24}$ and $C_{13}$ in the $b$-basis. In the $a$-basis those amplitudes are represented by
$$
\psi_{C_{24}}=\frac{1}{\sqrt{2}}\; [e_1^a - e_2^a], \;  \psi_{C_{13}}=\frac{1}{\sqrt{2}}\; [e_1^a + e_2^a]
$$

(d) Let $C=C_{123}=\{\omega_1, \omega_2, \omega_3\}.$ Here ${\bf P}(C)=(2q+1)/2,
{\bf P}(A_1/C)={\bf P}(B_2/C)=1/(2q+1), {\bf P}(A_2/C)={\bf P}(B_1/C)=2q/(2q+1).$
Thus $\delta(B_1/a, C)=\frac{2q(2q-1)}{2q+1}$ and,
hence, $\la(B_1/a, C)=-\frac{\sqrt{1-2q}}{2}.$ This context is trigonometric:
$C_{123}\in {\cal C}.$ We remark that $\la(B_2/a, C)=\frac{\sqrt{1-2q}}{2}$
(since ${\bf P}^{b/a}$ is double stochastic).\footnote{
We pay attention on the dependence of $\theta= \arccos \frac{\sqrt{1-2q}}{2}$
on the parameter $q:\theta(q)$ increases from $\pi/3$ to $\pi/2,$
when $q$ increases from 0 to 1/2.} We choose $\theta_2= \arccos \frac{\sqrt{1-2q}}{2},$ so
$\theta_1= \arccos\frac{\sqrt{1-2q}}{2} - \pi.$We have:
\[\psi_{C_{123}}(x)= \left\{ \begin{array}{ll}
{\sqrt{\frac{2q}{2q+1}}-e^{i \arccos \frac{\sqrt{1-2q}}{2}}\sqrt{\frac{2q(1-2q)}{2q+1}}, \;x=b_1}\\
{\sqrt{\frac{1-2q}{2q+1}}+e^{i \arccos \frac{\sqrt{1-2q}}{2}}{\frac{2q}{\sqrt{2q+1}}}, \;\;\;\;\;\;x=b_2}
\end{array}
\right. \]

Thus
 \[\psi_{C_{123}}=\frac{1}{\sqrt{2q+1}}e_1^a - ie^{i\arccos \frac{\sqrt{1-2q}}{2}}\sqrt{\frac{2q}{2q+1}}e_2^a\;.\]

\noindent
(e) Let $C=C_{124}=\{\omega_1, \omega_2, \omega_4\}.$
Here ${\bf P}(C)=1-q, {\bf P}(A_1/C)={\bf P}(B_1/C)=1/2(1-q), {\bf P}(A_2/C)={\bf P}(B_2/C)=(1-2q)/2(1-q).$
Thus $\delta(B_1/a, C)=q(1-2q)/(1-q)$ and, hence, $\la(B_1/a, C)=\sqrt{\frac{q}{2}}<1,$
and the context $C_{124} \in {\cal C}.$  We choose $\theta_1 = \arccos\sqrt{\frac{1}{2}},$
so $\theta_2 = \arccos\sqrt{\frac{1}{2}} + \pi.$ Thus:

\[\psi_{C_{124}}(x)= \left\{ \begin{array}{ll}
{\sqrt{\frac{q}{1-q}}+      e^{i \arccos \sqrt{\frac{q}{2}}}\frac{1-2q}{\sqrt{2(1-q)}},\;\;\;\;\; x=b_1}\\
{\sqrt{\frac{1-2q}{2(1-q)}} - e^{i \arccos \sqrt{\frac{q}{2}}}\sqrt{\frac{q(1-2q)}{1-q}},\; x=b_2}
\end{array}
\right . \]
\[\psi_{C_{124}}(x)=
\frac{1}{\sqrt{2(1-q)}} \;e_1^a + i e^{i \arccos \sqrt{\frac{q}{2}}}\sqrt{\frac{1-2q}{2(1-q)}}\; e_2^a.\]

(f) Let $C=C_{234}=\{\omega_2, \omega_3, \omega_4\}.$
Here ${\bf P}(C)=1-q, {\bf P}(A_1/C)={\bf P}(B_1/C)=(1-2q)/2(1-q), {\bf P}(A_2/C)={\bf P}(B_2/C)=1/2(1-q).$
Thus $\delta(B_1/a, C)=q(2q-1)/(1-q)$
and, hence, $\la(B_1/a, C)=-\sqrt{\frac{q}{2}},\la(B_2/a, C)=\sqrt{\frac{q}{2}}.$
Here:
\[\psi_{C_{234}}(x)= \left\{ \begin{array}{ll}
{\sqrt{\frac{q(1-2q)}{1-q}}-e^{i \arccos \sqrt{\frac{q}{2}}}\sqrt{\frac{1-2q}{2(1-q)}}, \;x=b_1}\\
{\frac{1-2q}{\sqrt{2(1-q)}}+e^{i \arccos \sqrt{\frac{q}{2}}}\sqrt{\frac{q}{1-q}}, \;\;\;\;\;\;x=b_2}
\end{array}
\right. \]
\[\psi_{C_{234}}(x)=\sqrt{\frac{1-2q}{2(1-q)}}\; e_1^a  - i e^{i \arccos \sqrt{\frac{q}{2}}}\frac{1}{\sqrt{2(1-q)}} \;e_2^a\;.\]

(g) Let $C=C_{134}=\{\omega_1, \omega_3, \omega_4\}.$
Here ${\bf P}(C)=(2q+1)/2, {\bf P}(A_1/C)={\bf P}(B_2/C)=2q/(2q+1), {\bf P}(A_2/C)={\bf P}(B_1/C)=1/(2q+1).$
Thus $\delta(B_1/a, C)=2q(1-2q)/(2q+1)$
and, hence, $\la(B_1/a, C)=\frac{\sqrt{1-2q}}{2}.$
Thus:
\[\psi_{C_{134}}(x)= \left\{ \begin{array}{ll}
      {\frac{2q}{\sqrt{2q+1}}+e^{i \arccos \frac{\sqrt{1-2q}}{2}}\sqrt{\frac{1-2q}{2q+1}}, \;\;\;\;\;x=b_1}\\
{\sqrt{\frac{2q(1-2q)}{2q+1}}-e^{i \arccos \frac{\sqrt{1-2q}}{2}}\sqrt{\frac{2q}{2q+1}}, x=b_2}
\end{array}
\right. \]

\[\psi_{C_{134}}=\sqrt{\frac{2q}{2q+1}} \;e_1^a + e^{i \arccos \frac{\sqrt{1-q}}{2}}\frac{1}{\sqrt{2q+1}} \;e_2^a.\]

\noindent
(h) Let $C=\Omega.$ Here we know from the beginning that $\delta(B_j/a, C)=0.$
Here ${\bf P}(A_i/C)={\bf P}(A_i)=1/2$ and ${\bf P}(B_i/C)={\bf P}(B_i)=1/2.$
We can choose the phase $\theta_\Omega(b_1)= \frac{\pi}{2}$ or $\frac{3\pi}{2}.$ The first
choice of gives the complex amplitude $\psi_\Omega= \psi_{C_{24}}$ and the second
$\psi_\Omega= \psi_{C_{13}}.$ Thus for both representations the map $J^{b/a}$ is
{\it not injective.}

The representation map $J^{b/a}$ transforms
the conventional probabilistic calculus in $\cal K$ to the quantum probabilistic calculus
in $H.$ For example, let a random variable $b=\pm 1.$ Here
$$
E(b/C_{234})={\bf P}(B_1/C_{234}) - {\bf P}(B_2/C_{234})=\frac{1-2q}{2(1-q)} - \frac{1}{2(1-q)} = \frac{q}{q-1}.
$$
Hence the conventional probabilistic calculus gives the answer $q/(q-1).$ But we also have:
\[<\hat b>_{\psi_{C_{234}}}=(\hat b \psi_{C_{234}}, \psi_{C_{234}})\]
\[= \left| \sqrt{\frac{q(1-2q)}{1-q}}-
e^{i \rm{arccos}  \sqrt{\frac{q}{2}}} \sqrt{\frac{1-2q}{2(1-q)}}\; \right|^2 \]
\[- \left| \sqrt{\frac{(1-2q)^2}{2(1-q)}} +
e^{i \rm{arccos} \sqrt{\frac{q}{2}}} \;\sqrt{\frac{q}{1-q}}\; \right|^2 \]
\[=\frac{q(1-2q)}{1-q} + \frac{1-2q}{2(1-q)} - 2 \sqrt{\frac{q}{2}\;\frac{q(1-2q)}{1-q} \;\frac{1-2q}{2(1-q)}}\]
\[-\frac{(1-2q)^2}{2(1-q)} - \frac{q}{1-q} - 2 \sqrt{\frac{q}{2}\;\frac{(1-2q)^2}{2(1-q)} \;\frac{q}{1-q}}=
\frac{q}{q-1}.\]
Thus the quantum probabilistic calculus gives us the same result $q/(q-1).$ In the same way
we have for a random variable  $a=\pm 1:$
$$
E(a/C_{234})={\bf P}(A_1/C_{234}) - {\bf P}(A_2/C_{234})=\frac{q}{q-1}
$$
$$
<\hat a>_{\psi_{C_{234}}}=(\hat a \psi_{C_{234}}, \psi_{C_{234}})=
\left| \sqrt{\frac{1-2q}{2(1-q)}}\; \right|^2 - \left| -ie^{ i \rm{arccos}  \sqrt{\frac{q}{2}}}
\frac{1}{\sqrt{2(1-q)}}\;\right|^2=\frac{q}{q-1}.
$$
In this example the set of nonsensitive contexts contains three contexts: ${\cal C}_0=
\{ \Omega, C_{24}, C_{13}\}.$ We have
$$
S_{\bar{\cal C}}= \{\psi_{C_{13}}, \psi_{C_{24}}, \psi_{C_{14}}=e_1^b, \psi_{C_{23}}=e_2^b,
\psi_{C_{12}}=e_1^a, \psi_{C_{23}}=e_2^a,
\psi_{C_{124}}, \psi_{C_{234}}, \psi_{C_{123}}, \psi_{C_{134}}\}
$$
Therefore the set of pure states $S_{\bar{\cal C}}$ is a finite, ten-points, subset of the unit sphere
in the two dimensional Hilbert space. We remark that all vectors in $S_{\bar{\cal C}}$ are pairwise
independent.

There is the parameter $q \in (0, 1/2)$ determining
a Kolmogorov probability space ${\cal K}(q).$ For each value of $q$ we have  a finite set of pure states.
However, a family ${\cal K}(q), q\in (0, 1/2),$ of Kolmogorov probability spaces generates a ``continuous'' set
$\cup_q S_{\bar{\cal C}}(q)$ of pure states.

Finally, we remark that we have chosen one fixed representation of every Kolmogorov model ${\cal K}(q)$ (for fixed $q).$
We can construct other representations corresponding to other choices of phases. For example, we can construct
another representation  by choosing
$
\theta_{C_{24}}(b_1) = \frac{3 \pi}{2}, \theta_{C_{13}}(b_1) = \frac{\pi}{2},
\theta_{C_{134}}(b_1) = - \arccos \frac{\sqrt{1-2q}}{2},$\\
$
\theta_{C_{123}}(b_2) = - \arccos \frac{\sqrt{1-2q}}{2},
\theta_{C_{124}}(b_1) = - \arccos \sqrt{\frac{q}{2}},
\theta_{C_{234}}(b_2) = - \arccos \sqrt{\frac{q}{2}},
$
as well as combine some phases choices of various representations.

\section{Hilbert space images of the reference random variables}

We begin with the following standard definition:

 {\bf Definition 3.} {\it For a self-adjoint operator $\hat d$ the quantum mean value in the state $\psi$ is defined by}
 \[\langle \hat d \rangle_\psi= (\hat d \psi, \psi).\]

 {\bf Theorem 5.} {\it{For any map $f:{\bf R} \to {\bf R},$ we have:}}
 \[\langle f(\hat a)\rangle_{\psi_C}=E(f(a)/C), \;\;\; \langle f(\hat b)\rangle_{\psi_C}=E(f(b)/C)\]
 for any context $C \in \bar {\cal C}.$

 {\bf Proof.} By using the Born's rule for the $b$ we obtain:
 \[E(f(b/C)=\sum_{x\in X} f(x) p_c^b(x)=\sum_{x \in X} f(x)|(\psi_C, e_x^b)|^2=\langle f(\hat b)\rangle_{\psi_C}\]
The same result we have for the $f(\hat a)$ since (as ${\bf P}^{b/a}$ is double stochastic)
we have Born's probability rule both for $b$ and $a.$

{\bf Theorem 6.} {\it Let $f, g: {\bf R} \to {\bf R}$ be two arbitrary functions.
Then
\[E(f(a) + g(b)/C)=\langle f(\hat a) + g(\hat b)\rangle_{\psi_C}\]
for any context $C \in {\bar {\cal C}}.$}

{\bf Proof.} By using linearity of the Kolmogorov mathematical expectation, Theorem 3,
and linearity of the Hilbert space scalar product we obtain:
\[E(f(a(\omega)) + g(b(\omega))/C)=E(f(a(\omega)/C) + E(g(b(\omega))/C)\]
\[= \langle f(\hat a)\rangle_{\psi_C} + \langle g(\hat b)\rangle_{\psi_C}=
\langle f(\hat a) + g(\hat b)\rangle_{\psi_C}\]

Denote the linear space of all random variables of the form
$d(\omega)=f(a(\omega)) + g(b(\omega))$ by the symbol ${\cal O}_+(a,b)$
and the linear space of operators of the form $\hat d=f(\hat a) + g(\hat b)$ by
${\cal O}_+(\hat a, \hat b).$

{\bf Theorem 7.} {\it  The map $T=T^{a/b}:{\cal O}_+(a, b) \to {\cal O}_+(\hat a, \hat b),
d=f(a) + g(b) \to \hat d=f(\hat a) + g(\hat b),$
preserves the conditional expectation:}
\begin{equation}
\label{R}
\langle T(d)\rangle_{\psi_C}=(T(d) J(C), J(C))=E(d/C).
\end{equation}

The transformation $T$ preserves the conditional expectation for random variables $d \in {\cal O}_+(a, b).$
  But in general we cannot expect anything more, since in general $T$ does not preserve
  probability distributions. The important problem is to
  extend the map $T$ for a larger class (linear space?) of Kolmogorovian random variables
  with preserving (\ref{R}). It is natural  to define (as we always do in the conventional quantum formalism):
  $$
  T(f)(\hat a, \hat b)=f(\hat a, \hat b)
  $$
  where $f(\hat a, \hat b)$ is the pseudo differential operator with the Weyl symbol $f(a, b)$.
  We shall see that already for $f(a, b)=ab$ (so $f(\hat a, \hat b)=(\hat a \hat b + \hat b \hat a)/2$)
  the equality (\ref{R}) is violated.

We can consider the $b$ and the $a$ as discrete analogues of the position and momentum observables.
The operators $\hat b$ and $\hat a$ give the Hilbert space (quantum) representation of these observables.
We also introduce an analogue of the energy observable:
$$
{\cal H}(\omega)=\frac{a^2(\omega)}{2m} + V(b(\omega)),
$$
where $V:{\bf R} \to {\bf R}$ is a map.
The Hilbert space representation of this observable is given by the operator of energy (Hamiltonian)
$$
\hat {\cal H} =\frac{\hat a^2}{ 2m} + V(\hat b).
$$
By Theorem 7 for contexts $C \in \bar{\cal C}$ the averages of the observables ${\cal H}(\omega)$
(Kolmogorovian) and $\hat {\cal H}$ (quantum) coincide:
$$
E({\cal H} (\omega)/C) = \langle {\cal H} \rangle_{\psi_C} .
$$
However, as we shall see, probability distributions do not coincide:

{\bf Proposition 4.} {\it There exists  context $C$ such that the probability distribution
of the random  variable $d(\omega)=a(\omega) + b(\omega)$ with respect to $C$ does not
coincide with the probability distribution of the quantum observable
$\hat d=\hat a + \hat b$ with respect to the state $\psi_C$ corresponding to $C.$}

{\bf Proof.} It suffices to present an example of such a context $C.$
Take the context $C=C_{234}$ from the Example. We consider the case:
$a(\omega)=\pm \gamma, b(\omega)=\pm \gamma, \gamma > 0;$ so
$d(\omega)= - 2\gamma, 0, 2\gamma.$
Corresponding Kolmogorovian probabilities can easily be found:
\[p_C^d(-2\gamma)= q/(1-q),\; \;  p_C^d(0)=(1-2q)/(1-q),\; \;  p_C^d(2\gamma)=0.\]
We now find the probability distribution of $\hat d$.
To do this, we find eigenvalues and eigenvectors of the self-adjoint operator
$\hat d.$
We find the matrix of the operator $\hat d$ in the basis
$\{ e_j^b \}:$
$ d_{11} = - d_{22}= 4q \gamma $ and $d_{12} = d_{21}=2 \gamma \sqrt{2q(1-2q)}.$
We have $k_{1,2}=\pm 2 \sqrt{2q}\gamma.$
Of course, the range of
values of the quantum observable $\hat d$ differs from the range of values of the random variable $d.$
However, this difference of ranges of values is not so large problem in this case. The random variable $d$
takes only two values, $-2\gamma, 0$ with the probability one. Moreover, we can represent values of
the quantum observable $\hat d$ as just an affine transform of values of the random variable $d:$
\[d_{\rm{quantum}}=2\sqrt{2q} \; d-\gamma.\]
In principle we can interpret such a transformation as representing some special measurement procedure.
Thus in this example the problem with spectrum is not crucial. The crucial problem is that $d$ and
$\hat d$ have different probability distributions.

Corresponding eigenvectors are
\[e_1^d=\frac{1}{\sqrt{2(1-\sqrt{2q})}} (-\sqrt{1-2q}, \sqrt{2q}-1)\]
\[e_2^d=\frac{1}{\sqrt{2(1+\sqrt{2q})}} (-\sqrt{1-2q}, \sqrt{2q}+1)\]

Finally, we find (by using the expression for $\psi_{C_{234}}$ which was found in section 6):
\[p_c^{\hat d}(k_1)=|(\psi_C, e_1^d)|^2=\frac{(1-\sqrt{2q})(2+\sqrt{2q})}{4(1-q)}\]
\[p_c^{\hat d}(k_2)=|(\psi_C, e_2^d)|^2=\frac{(1+\sqrt{2q})(2-\sqrt{2q})}{4(1-q)}\]
Thus $d$ and $\hat d$ have essentially different probability distributions.

\section{Dispersion-free states}
As originally stated by von Neumann, [5] the problem of hidden variables is to find whether
{\it{dispersion free states exist}} in quantum mechanics. He answered the question in the negative.
The problem of the existence of dispersion free states as well as von Neumann's solution were
the subject of great debates. We do not want to go into detail see, e.g., [11]. In our contextual
approach an analogue of this problem can be formulated as: {\it Do dispersion free contexts exist?}

The answer is the positive. In the Example  we can take any atom of
the Kolmogorov probability space ${\cal K}_q,$ e.g., $C=\{\omega_1\}.$ Since, for any random variable
$\xi$ on the Kolmogorov space ${\cal K}_q$, it has a constant value on such a $C$
the dispersion of $\xi$ under the context $C$ is equal to zero:

\[D(\xi/C)=E[(\xi-E(\xi/C))^2/C]=0.\]

However, dispersion free contexts do not belong to the system $\bar{\cal C}$ of contexts which
can be mapped by $J^{a/b}$ into the Hilbert space $H.$ On the one hand, our contextual approach gives
the possibility to have the realist viewpoint to quantum mechanics..  On the other hand, it does not
contradict to the
von Neumann as well as other ``no-go" theorems. The mathematical representation of contexts
(complexes of physical conditions) given by the quantum formalism it too rough to represent dispersion
free contexts.

{\bf Conclusion:} {\it Dispersion free contexts exist, but they could not be represented by quantum states
(complex probabilistic amplitudes).}

\section{Classical and quantum spaces as rough images of fundamental prespace}

Our contextual probabilistic model induces the following picture of physical reality.

\subsection{Prespace and classical space}

There exists a prespace $\Omega$ which points corresponds to primary (irreducible) states
of physical systems, {\bf prestates or fundamental physical parameters}. Functions
$d:\Omega \to {\bf R}^m$ are said to be {\bf preobservables.} The set of all preobservables
is denoted by the symbol ${\cal O}_p \equiv {\cal O}_p (\Omega).$ We are not able (at least at the moment)
to measure an arbitrary preobservable  $d \in {\cal O}_p.$

Nevertheless, some preobservables can be measured. Suppose that there exists a preobservable
$b$ such that all measurements can be reduced to some measurements of $b,$ cf. L. De Broglie [12] and
D. Bohm [8].
on the possibility to reduce any measurement to a position measurement. Let $X \subset {\bf R}^m$ be the
range of values of $b.$ The $X$ is said to be a classical space \footnote{ Of course, in such a model
the classical space $X$ depends on the preobservable $X\equiv X(b).$ Thus $X$ is the $b$-image of
the prespace $\Omega.$}. Set $B_x=\{\omega \in \Omega: b(\omega)=x\}=b^{-1}(x), x \in X.$

In principle a set $B_x$ could contain millions of points. Dynamics in $X$ is classical dynamics.
 In our model, classical dynamics is a rough image of dynamics in the prespace $\Omega$
  \footnote{ Consider in the Example the trajectory
  $\omega_1 \to \omega_2 \to \omega_3 \to \omega_4 \to \omega_1$ in the $\Omega$.
  In the classical space $X$ this trajectory is represented by $b_1 \to b_1 \to b_2 \to b_2 \to b_1$.}.

\subsection{Classical phase space}
Let $a$ be a preobservable which is {\it incompatible} with our fundamental preobservable $b$
(space observable). We denote by $Y \subset {\bf R}^m$ the range of values of the $a.$ The $Y$
is said to be conjugate space to the classical space $X.$ We call the $b$ {\it{position}}
and the $a$ {\it{momentum.}} We set $A_y=\{\omega \in \Omega: a(\omega)=y\}=a^{-1}(y), y \in Y.$

Since $A_y$ is not a subset of $B_x$ for any $x \in X$ (this is a consequence of incompatibility of the
observables $a$ and $b),$  the point $y$ cannot
be used to get finer description of any point $x \in X.$ Thus by using
values of $a$ we cannot obtain a finer space structure. The variables $b$ and $a$ are really
incompatible. By fixing the value of, e.g., $a=y_0$ we cannot fix the value of $b=x_0.$

{\bf Remark.}
(Nonlocal dependence of incompatible variables at the prespace level). Since, for a fixed $y_0 \in Y,$
we have $A_{y_0} \cap B_x \ne \emptyset$ for any $x \in X,$ a value $y_0$ of the momentum can
be determined only by all values $x \in X$ of the position. Thus on the level of
the prespace incompatible variables are {\bf{nonlocally dependent}}.
However, this prespace nonlocality could not be found in classical mechanics, since in the latter
the finer prespace structure is destroyed by the rough $(x, y)$ encoding.

The space $\Pi=X\times Y \subset {\bf R}^{2m}$ is a classical phase space. Dynamics in the phase space gives
a rough image in the terms of the two incompatible variables of dynamics in the prespace.
The phase space $\Pi$ is a classical contextual $(b, a)$-picture of the prespace $\Omega.$ This picture is
richer than the pure $b$-space picture,$X.$ The $\Pi$ contains images of the
two families of contexts ${\cal A}=\{A_y\}$ and ${\cal B}=\{B_x\}.$

In our probabilistic investigations we have seen that the most
natural choice of incompatible variables corresponds to random variables $a(\omega)$ and
$b(\omega)$ which are uniformly distributed. On the other hand, the creation of a uniform partition
of the prespace $\Omega$ is the most natural way to create a rough image $X$ of the prespace -- a classical space.

As the next step we can consider statistical mechanics on the classical space $X.$ In such a statistical
theory from the very beginning we lost the finer statistical structure of the model based on probability
distributions on the prespace. Functions $u: \Pi \to {\bf R}^q$ are called classical observables.
The set of classical observables is denoted by the symbol ${\cal O}_c (\Pi).$
We shall also use symbols ${\cal O}_c (X)$ and  ${\cal O}_c (Y)$ to denote spaces of classical observables
depending only on the $b$-position and the $a$-momentum, respectively.

\subsection{Quantum mechanics and the Hilbert space representation of prespace contexts}
Neither classical nor quantum mechanics can describe the individual dynamics in the
prespace. Of course, such a viewpoint to quantum mechanics contradicts to the so called orthodox
Copenhagen interpretation by which the wave function describes an individual quantum system.
It also contradicts to the conventional viewpoint to classical mechanics.

By our {\it contextual interpretation} the wave function has the realist prespace interpretation.
A complex amplitude is nothing than the  image (induced by the contextual formula of total
probability) of a set of fundamental parameters - a context. Thus the Hilbert state space $H$
is not less real than the classical real space ${\bf R}^3.$

Observables which probability distributions can be found by using the representation by self-adjoint
operators in the Hilbert space are called quantum observables. The set of quantum observables is denoted
by the symbol ${\cal O}_q(H).$
Neither classical statistical nor quantum mechanics can provide knowledge about the probability distribution of an arbitrary preobservable. Nevertheless, the quantum theory gives some information about some preobservables, namely
fundamental preobservable $b$ and $a$ and pre-observables $d$ belonging to the class ${\cal O}_+ (a, b).$
Another way to look to the same problem is to say that the quantum
theory (with our contextual probabilistic interpretation) gives the possibility to represent
some prespace structures, namely some contexts $C \in {\cal C}$ by vectors of a Hilbert state space.

Neither classical nor quantum mechanics are fundamental theories. They could not give
information about the point wise structure of the prespace $\Omega.$ But the quantum formalism
represents some complexes of physical conditions -- domains in the prespace -- which are not represented in
the classical space or phase space. Of course, the quantum formalism also represents classical position
states $x \in X$ by wave functions $\psi_{B_x}$ (Hilbert states $e_x^b$). Classical states $x \in X$ are
images of prespace contexts $B_x.$ But the quantum formalism represents also some sets $C \subset \Omega$
which have no classical images (namely, images in $X$ or $\Pi$).

In the Example  we take the set $C=C_{123}=\{\omega_1, \omega_2, \omega_3\}.$
Neither $C \subset B_1$ nor $C \subset B_2.$ This prespace domain $C$ can be described neither by
the position $x=b_1$ nor $x=b_2$. The quantum state $\psi_C \in S \subset H$ representing this
domain of the prespace describes the superposition of the two classical states $x=b_1$ and $x=b_2.$
Hence a physical system prepared under the complex physical conditions $C=C_{123}$ is (from the classical
viewpoint) in the superposition of two different positions.

\subsection{Heisenberg uncertainty principle}
We now take the context $C=A_y$ for some $y \in Y.$
Here the momentum $a$ has the definite value. But  $A_y \cap B_x \ne \emptyset$ for any $x \in X.$
Hence the state $\psi_C=e_y^a \in H$ also corresponds to the superposition of
two positions  $x=b_1$ and $x=b_2.$
This is nothing else than (the discrete analogue) the {\bf Heisenberg uncertainty
principle}. In the same way in any state with the definite position, $\psi_C=e_x^b, x\in X,$ the
momentum can not have the definite value.

Thus the Hilbert sphere $S$ contains images of the classical spaces $X$ and $Y$
(but not the phase space  $\Pi$, see further considerations), $X\equiv \{e_x^b\} \subset S$ and $Y=\equiv\{e_y^a\} \subset S.$ But the Hilbert
space contains also images of nonclassical domains
$C \in \bar{\cal C}.$ We remark that (depending on the model) only a part of the Hilbert sphere
corresponds to some domains of the prespace. All other quantum states, $\psi \not\in S_{\bar {\cal C}},$
are just ideal mathematical objects which do no correspond to any context in the prespace.

As was already remarked, the phase space  $\Pi$
is not imbedded into the Hilbert sphere $S$, since contexts $C_{xy}=B_x \cap A_y$
corresponding to points of the $\Pi$ do not belong to the system $\bar {\cal C}$ which is mapped
into $S$ (because these contexts are degenerate with respect to both reference observables).

\subsection{ Preobservables and quantum observables}
For what class of preobservables can we find probability distributions with respect to contexts
$C \in \bar{\cal C}$ by using the quantum formalism? As we have seen, we are not able to find the
probability distribution for an arbitrary preobservable $d\in {\cal O}_p(\Omega).$
In general the operators $\hat d=d(\hat a, \hat b)$ corresponding to functions
$d(x, y)$ (e.g., $d(x, y)=xy$ or $d(x, y)=x+y)$
are not directly related to prequantum observables $d(\omega)=d(b(\omega), a(\omega)).$

Only quantum observables $\hat d=f(\hat b)$ and $\hat d=g (\hat a)$ have the same probability
distributions as the corresponding preobservables $d(\omega)=f(b(\omega))$ and $d(\omega)=g(a(\omega)).$
By Theorem 7 the average is preserved by the canonical map $T^{b/a}: {\cal O}_+(a, b) \to
{\cal O}_+(\hat a, \hat b).$

However, even such quantum observables give just a rough image of corresponding preobservables.
By using quantum probabilistic formalism we can find probability distributions only for quantum
states $\psi_C \in S_{\bar{\cal C}} \subset H$.
Those quantum states represent only some special contexts. Hence by using the quantum formalism
we could not find the probability distribution of a preobservable $a(\omega)$ or $b(\omega)$ for
an arbitrary context represented by a domain in the prespace $\Omega.$ Neither we can reconstruct maps
$a(\omega)$ and $b(\omega)$. Thus the quantum theory is not a fundamental theory. It does not provide
the complete (even statistical) description of the prespace reality. However, some statistical information
about the prespace structure can be obtained by using the quantum probabilistic formalism.

\subsection{On the mystery of  operator quantization}
The origin of the operator quantization was always mysterious for me. Why the correspondence between functions and functions of operators (of the position and the momentum) provides the correct statistical description of quantum measurements? Our contextual model tells that the only reason is the coincidence of quantum averages with `real' prespace (contextual) averages for some preobservables (in particular, of the form $f(b)+ g(a)).$

Theorem 7 is only a sufficient condition for the coincidence of averages.
But even such a result gives
the possibility to connect the quantum Hamiltonian
$$\hat {\cal H} =\frac{\hat a^2}{2m} + V(\hat b)$$
with the realist preobservable ${\cal H}(a(\omega), b(\omega))=\frac{a(\omega)^2}{2m} + V(b(\omega)).$
Quantum averages of energy expressed by the Hilbert space averages of the Hamiltonian
$\hat {\cal H}$ coincides with averages of the realist energy preobservable ${\cal H}(\omega).$
However, for some contexts $C$ quantum energy observable $\hat {\cal H}$ and energy
preobservable ${\cal H}$
have different probability distributions, see Proposition 4.
 In principle, such an effect should be observable
experimentally.

The classical space is a contextual image of the fundamental prespace $\Omega.$
 This is a very poor image since only a few special contexts namely space-contexts have images
 in the classical space ${\bf R}^3$. In principle, there might be created
 various classical spaces (corresponding to various fundamental variables)
 on the basis of the prespace $\Omega$. Human beings have been creating their own (very special)
 classical space. Since light rays play the fundamental role in the creating of our classical
 space it can be called {\it{electromagnetic classical space}}. So  the electromagnetic classical space
 is created on the basis on electromagnetic reduction of information. In principle there can exist
 systems which are able to perform some other reductions of information, e.g., gravitation reduction.
 They would create a {\it gravitational classical space.}

\section{Hyperbolic Hilbert space projection of the classical probabilistic model}

We study here the model with the {\it hyperbolic interference.} We set
$$
{\cal C}^{\rm hyp}=\{C\in{\cal C}_{a}:|\la(B_j/a, c)|\geq 1, j=1,2\}.
$$
We call elements of ${\cal C}^{\rm hyp}$ hyperbolic contexts.

\subsection{Hyperbolic algebra}
Instead of the field complex numbers
${\bf C},$  we shall use  so called {\bf hyperbolic numbers,} namely the two dimensional
Clifford algebra, ${\bf G},$ see [49]. We call this algebra {\it hyperbolic algebra.}
\footnote{Of course, it is rather dangerous to invent an own name for a notion established almost as
firm as complex numbers. We use a new name,  hyperbolic algebra, for the well known algebraic
object, the two dimensional Clifford algebra, by following reasons. First we explain why we dislike
to use the standard notion Clifford algebra in this particular case. The standard Clifford machinery
was developed around noncommutative features of general Clifford algebras. The two dimensional Clifford
algebra, hyperbolic algebra in our terminology, is commutative. Commutativity of ${\bf G}$ is very important
in our considerations. We now explain why we propose the name hyperbolic algebra.
Hyperbolic functions are naturally related to the algebraic structure
of ${\bf G}$ through a hyperbolic generalization of Euler's formula for the complex
numbers. This is the crucial point of our considerations - the possibility to use this
algebraic structure to represent some special transformations for hyperbolic functions.}

Denote by the symbol $j$  the generator of the algebra ${\bf G}$ of hyperbolic numbers:
$$
j^2=1.
$$
The algebra {\bf{G}} is the two dimensional real
algebra with basis $e_0=1$ and $e_1=j.$  Elements of {\bf{G}} have the form $z=x + j y, \; x, y \in {\bf{R}}.$
We have $z_1 + z_2=(x_1+x_2)+j(y_1+y_2)$ and $z_1 z_2=(x_1x_2+y_1y_2)+j(x_1y_2+x_2y_1).$
This algebra is commutative. It is not a field - not every element has the inverse one.

We introduce an involution in {\bf{G}} by setting
$\bar{z} = x - j y$
and set  $|z|^2=z\bar{z}=x^2-y^2.$
We remark that  $|z|=\sqrt{x^2-y^2}$ {\bf is not well defined} for an arbitrary $z\in {{\bf{G}}}.$
We set ${{\bf{G}}}_+=
\{z\in{{\bf{G}}}:|z|^2\geq 0\}.$ We remark that ${{\bf{G}}}_+$
is a multiplicative semigroup as follows from the equality

$|z_1 z_2|^2=|z_1|^2 |z_2|^2.$

Thus, for $z_1, z_2 \in {{\bf{G}}}_+,$
we have that $|z_1 z_2|$ is well defined and
$|z_1 z_2|=|z_1||z_2|.$ We define a hyperbolic exponential function by using
a hyperbolic analogue of the Euler's formula:

$e^{j\theta}=\cosh\theta+ j \sinh\theta, \; \theta \in {\bf{R}}.$

We remark that

$e^{j\theta_1} e^{j\theta_2}=e^{j(\theta_1+\theta_2)}, \overline{e^{j\theta}}
=e^{-j\theta}, |e^{j\theta}|^2= \cosh^2\theta - \sinh^2\theta=1.$

Hence, $z=\pm e^{j\theta}$ always belongs to ${{\bf{G}}}_+.$
We also have

$\cosh\theta=\frac{e^{j\theta}+e^{-j\theta}}{2}, \;\;\sinh\theta=\frac{e^{j\theta}-e^{-j\theta}}{2 j}\;.$

We set ${{\bf{G}}}_+^*=
\{z\in{{\bf{G}}}_+:|z|^2>0 \}. $
Let  $z\in {{\bf{G}}}_+^*.$  We have

$z=|z|(\frac{x}{|z|}+j \frac{y}{|z|})= \rm{sign}\; x\; |z|\;(\frac{x {\rm{sign}} x}{|z|} +j\;
\frac{y {\rm{sign}} x}{|z|}).$

As $\frac{x^2}{|z|^2}-\frac{y^2}{|z|^2}=1,$  we can represent $x$ sign $x= \cosh\theta$
and $y$ sign $x=\sinh\theta, $ where the phase $\theta$ is unequally defined.
We can represent each $z\in {{\bf{G}}}_+^*$ as

$z = \rm{sign}\; x\;  |z|\; e^{j\theta}\;.$

By using this representation we can easily prove that ${{\bf{G}}}_+^*$
is a multiplicative group. Here $\frac{1}{z}=\frac{{\rm{sign}} x}{|z|}e^{-j\theta}.$
The unit circle in ${{\bf{G}}}$ is defined as $S_1 = \{z\in{{\bf{G}}}:|z|^2=1\}
=\{ z= \pm e^{j \theta}, \theta \in (-\infty, +\infty)\}.$ It is a multiplicative
subgroup of ${\bf G}_+^*.$

\subsection{Hyperbolic probability amplitude, hyperbolic Born's rule}
The interference formula of total probability (\ref{TNC1}) can be written in the following form:
\begin{equation}
\label{TwoH}
p_C^b(x)=\sum_{y \in Y}p_C^a(y) p(x/y) \pm 2\cosh \theta_C(x)\sqrt{\Pi_{y \in
Y}p_C^a(y) p(x/y)}\;,
\end{equation}
where $\theta_C(x)=\theta(x/a, C) = \pm \rm{arccosh} \vert \lambda(x/a, C)\vert,
x \in X, C \in {\cal C}^{\rm hyp}.$ Here the coefficient $\lambda$ is defined by (\ref{TNCT2}).
By using the elementary formula
\[D=A+B\pm 2AB\cosh \theta=|\sqrt{A}\pm e^{j\theta}\sqrt{B}|^2,\]
for $A, B>0,$ we can represent the probability $p_C^b(x)$ as the square of the
hyperbolic amplitude $p_C^b(x)=|\psi_C(x)|^2,$
where
\begin{equation}
\label{TwoA}
\psi(x) \equiv \psi_C(x)= \sqrt{p_C^a(a_1)p(x/a_1)} + \epsilon_C(x)
e^{j\theta_C(x)} \sqrt{p_C^a(a_2)p(x/a_2)} \;.
\end{equation}
Here $\epsilon_C(x)={\rm sign}\; \delta(x/a, C).$ We remark that by Lemma 1:
\begin{equation}
\label{MART}
\sum_{x\in X}\epsilon_C(x)=0.
\end{equation}
Thus we have a {\it hyperbolic generalization of Born's rule} for the $b$-variable.

\subsection{Hyperbolic Hilbert space representation}

{\bf  Definition 4.} {\it  A hyperbolic Hilbert space is
${{\bf{G}}}$-linear space (module) $H$
with a ${{\bf{G}}}$-linear scalar
product: a map $(\cdot, \cdot): H\times H \to {{\bf{G}}}$ that is

1) linear with respect to the first argument:

$ (a z+ b w, u) = a (z,u) + b (w, u), a,b \in {{\bf{G}}},
z,w, u \in H;$

2) symmetric: $(z,u)= \overline{(u,z)} ;$

3) nondegenerate: $(z,u)=0$ for all $u \in H$ iff $z=0.$}

{\bf Remark.} If we consider $H$ as just a ${\bf R}$-linear space, then $(\cdot, \cdot)$
is a bilinear form which is not positive defined.
In particular, in the two dimensional case we have the signature: $(+,-,+,-).$

 We introduce on the space
$\Phi(X, {\bf G})$ of functions: $\psi: X\to {\bf G}.$
Since $X= \{b_1, b_2 \},$ the $\Phi(X, {\bf G})$ is the two dimensional ${\bf G}$-module.
We define the ${\bf G}$-scalar product by (\ref{BHS}) with conjugation in
${\bf G}.$
The system of functions $\{e_x^b\}_{x\in X}$ is an orthonormal basis in the hyperbolic
Hilbert space $H^{\rm hyp}=(\Phi(X, {\bf G}), (\cdot, \cdot)).$
Thus we have the hyperbolic Born's rule in $H^{\rm hyp},$  see (\ref{BH}) -- but with the
hyperbolic scalar product.
The random variable $b$ is represented by the multiplication operator $\hat b$ in $\Phi(X, {\bf G}).$
We have the hyperbolic Hilbert space representation (\ref{BI1}) of the average of $b.$

Thus we constructed a ${\bf G}$-linear representation of the contextual Kolmogorov model:
$$
J^{b/a}: {\cal C}^{\rm{hyp}} \to H^{\rm hyp}.
$$
We set $S_{{\cal C}^{\rm{hyp}}} = J^{b/a} ({\cal C}^{\rm{hyp}} ).$ This is a subset of the unit sphere
$S$ of the Hilbert space $H^{\rm hyp}.$

By introducing the coefficients (\ref{KOE}) and $\epsilon_i=\epsilon(b_i)$ we represent a
state $\psi_C$ by $\psi_C=v_1^b e_1^b + v_2^b e_2^b,$
where $v_i^b=u_1^a u_{1i} + \epsilon_i u_2^a u_{2i} e^{j\theta_i}.$
So
$$
p_C^b(b_i)=|v_i^b|^2=|u_1^a u_{1i} + \epsilon_i u_2^a u_{2i} e^{j\theta_i}|^2 \;.
$$
This is the {\it G-linear representation of the hyperbolic interference of probabilities.}
This formula can also be derived in the formalism of the hyperbolic Hilbert space.
We remark that here the ${\bf G}$-linear combination $u_1^a u_{1i} + \epsilon_i u_2^a u_{2i} e^{j\theta_i}$
belongs to ${{\bf{G}}}_+^*.$

Thus for any context $C_0\in{\cal C}^{\rm hyp}$ we can represent $\psi_{C_0}$ in the form:
\[\psi_{C_0}=u_1^a e_1^a + u_2^a e_2^a,\] where
$$
e_1^a=(u_{11}, u_{12}) \; ,
e_2^a=(\epsilon_1 e^{j\theta_1} u_{21}, \epsilon_2 e^{j\theta_2} u_{22}).
$$
As in the $\bf C$-case, we introduce the matrix $V$ with coefficients
$v_{11}=u_{11}, v_{21}=u_{21}$ and $v_{12}=\epsilon_1e^{j\theta_1} u_{21},
v_{22}=\epsilon_2e^{j\theta_2} u_{22}.$ We remark that here coefficients $v_{ij} \in {{\bf{G}}}_+^*.$
In the same way as in the complex case the Born's rule
\begin{equation}
\label{ME}
p_{C_0}^a (a_i)=|(\psi_{C_0}, e_i^a)|^2
\end{equation}
holds true in the $a$-basis iff $\{e_i^a\}$ is an orthonormal basis in $H^{\rm hyp}.$
The latter is equivalent to the $\bf G$-unitary of the matrix $V$
(corresponding to the transition from $\{e_i^b\}$ to $\{e_i^a\}): \overline{V}^*V=I,$ or
\begin{equation}
\label{NORM1}
\bar v_{11} v_{11} + \bar v_{21} v_{21}=1,\;
\bar v_{12} v_{12} + \bar v_{22} v_{22}=1,
\end{equation}
\begin{equation}
\label{NORM2}
\bar v_{11} v_{12} + \bar v_{21} v_{22}=0.
\end{equation}
Thus $1=u_{11}^2 + u_{21}^2=p(b_1/a_1) + p(b_1/a_2)$ and $1=u_{12}^2 + u_{22}^2=p(b_2/a_1) + p(b_2/a_2).$
Thus the first two equations of the $\bf G$-unitary are equivalent to the double stochasticity
of ${\bf P}^{b/a}$ (as in the ${\bf C}$-case). We remark that the equations (\ref{NORM1}) can be
written as
\begin{equation}
\label{NORM3}
\vert v_{11}\vert^2 + \vert v_{21}\vert^2 =1, \vert v_{12}\vert^2 + \vert v_{22}\vert^2=1 .
\end{equation}
The third unitarity equation (\ref{NORM2}) can be written as
\begin{equation}
\label{NORM4}
u_{11} u_{12} \epsilon_1 e^{-j\theta_2} + u_{21} \epsilon_2 e^{-j\theta_2} u_{22}=0.
\end{equation}
By using double
stochasticity of ${\bf P}^{a/b} $ we obtain
$e^{j\theta_1}=e^{j\theta_2}.$ Thus
\begin{equation}
\label{MARE}
\theta_1=\theta_2.
\end{equation}
This is the hyperbolic analogue of the $\bf C$-unitary condition (\ref{MAR}).

{\bf Lemma 6.}
{\it{Let $a$ and $b$ be incompatible random variables and let ${\bf P}^{b/a}$ be double stochastic. Then
\begin{equation}
\label{Ka}
\cosh \theta_C(b_2)=\cosh \theta_C(b_1)
\end{equation}
for any context $C\in{\cal C}^{\rm hyp}$.}}

{\bf Proof.} By Lemma 1 we have:
\[\sum_x\epsilon(x)\cosh \theta_C(x)\sqrt{\Pi_y p_C^a(y) p(x/y)} =0.\]
Double stochasticity of ${\bf P}^{b/a}$ implies (\ref{Ka}).

\medskip

The constraint (\ref{Ka}) induced by double stochasticity
can be written as the constraint to phases:
\begin{equation}
\label{MK}
\theta_C(b_2)=\pm \theta_C(b_1).
\end{equation}
To obtain unitary of the matrix $V$ of transition $\{e_i^b\}\to \{e_i^a\}$ we should choose
phases according to (\ref{MARE}). And by (\ref{MK}) we can always do this for a double stochastic matrix of transition
probabilities.

By choosing such a representation we obtain the
hyperbolic generalization of the Born's rule (\ref{BBR}) for the $a$-variable.

We now investigate the possibility to use one fixed
basis $\{e_j^a \equiv e_j^a(C_0)\}, C_0 \in {\cal C}^{\rm hyp},$ for all
states $\psi_C, C\in {\cal C}^{\rm hyp}.$ For any $C\in {\cal C}^{\rm hyp}$ we
would like to have the representation (\ref{LUU}). We have
$$\psi_C(b_1)=u_1^a(C) v_{11}(C_0) +
\epsilon_C(b_1) \epsilon_{C_0}(b_1) e^{j[\theta_C(b_1) - \theta_{C_0}(b_1)]} u_2^a(C) v_{12}(C_0)$$
$$\psi_C(b_2)=u_1^a(C) v_{21}(C_0) +
\epsilon_C(b_2) \epsilon_{C_0}(b_2) e^{j[\theta_C(b_2) - \theta_{C_0}(b_2)]} u_2^a(C) v_{22}(C_0)$$

Thus to obtain (\ref{LUU}) we should have
\[\epsilon_C(b_1) \epsilon_{C_0}(b_1) e^{j [ \theta_C(b_1) - \theta_{C_0}(b_1)]}=
\epsilon_C(b_2) \epsilon_{C_0}(b_2) e^{j [\theta_C(b_2) - \theta_{C_0}(b_2)]}\]
Thus
\[\;\; \theta_C(b_1)-\theta_{C_0}(b_1)=\theta_C(b_2)-\theta_{C_0}(b_2), \; \rm{or}
\;\; \theta_C(b_1)-\theta_{C}(b_2)=\theta_{C_0}(b_1)-\theta_{C_0}(b_2).\]
By choosing the representation
with (\ref{MARE}) we satisfy the above condition.

\medskip

{\bf Theorem 8.}  {\it We can construct the hyperbolic Hilbert space representation of
the contextual Kolmogorov probability model such that the hyperbolic Born's rule holds true for both reference
variables $a$ and $b$ iff the matrix of transition probabilities ${\bf P}^{b/a}$ is double stochastic.}

We remark that by Theorem 5 basic
contexts $B_x, x \in  X,$ always belong to ${\cal C}^{\rm hyp},$ so $\psi_{B_x}\in H^{\rm hyp};$ and
$B_x\in {\cal C}^{\rm tr} \cap {\cal C}^{\rm hyp}$ iff $a$ and $b$ are uniformly distributed
(${\bf P}^{a/b}$ and ${\bf P}^{b/a}$ are double stochastic).

\subsection{Hyperbolic quantum mechanics}
As in the ordinary quantum formalism,
we represent physical states by normalized vectors of a hyperbolic Hilbert space $H:$
$\psi\in H$ and $(\psi, \psi)=1.$
We shall consider only dichotomous physical variables and quantum states belonging to
the two dimensional Hilbert space. Thus everywhere below $H$ denotes the two dimensional
space. Let $a=a_1, a_2$ and $b=b_1, b_2$ be two
physical variables. We represent they by  ${{\bf{G}}}$-linear operators:
$\hat{a}= \vert a_1> < a_1\vert + \vert a_2> < a_2\vert$ and $\hat{b}=
\vert b_1> < b_1\vert + \vert b_2> < b_2\vert,$
where $\{\vert a_i>\}_{i=1,2}$ and  $\{\vert b_i>\}_{i=1,2}$ are two orthonormal bases
in $H.$ The latter condition plays the fundamental role in hyperbolic quantum mechanics.
This is an analogue of the representation of physical observables by self-adjoint operators
in the conventional quantum mechanics (in the complex Hilbert space).

Let $\psi$ be a state (normalized vector belonging to $H).$ We can perform
the following operation (which is well defined from the mathematical point of view).
We expend the vector $\psi$ with respect
to the basis\footnote{We remark that we consider the two dimensional ${\bf G}$-Hilbert
space. There exists (by definition) a basis consisting of two vectors.}
$\{\vert b_i>\}_{i=1,2}:$
\begin{equation}
\label{E1}
\psi = v_1^b \vert b_1>+ v_2^b \vert b_2>,
\end{equation}
where the coefficients (coordinates) $v_i^b$ belong to ${\bf G}.$
As the basis $\{\vert b_i>\}_{i=1,2}$ is orthonormal, we have (as in the complex case) that:
\begin{equation}
\label{E2}
|v_1^b|^2 + |v_2^b|^2 = 1\;.
\end{equation}
However, we could not automatically use Born's probabilistic interpretation for
normalized vectors in the hyperbolic Hilbert space:
it may be that $v_i^b \not\in {\bf G}_+$ \footnote{In fact, in the complex case we have
${\bf C}={\bf C}_+$; thus there is no problem with positivity.} and hence $\vert v_i^b \vert^2 < 0.$
Since we do not want to consider negative probabilities (cf. [50]), in such a case we cannot use
the hyperbolic version of Born's probability interpretation.

{\bf Definition 5.} {\it A state $\psi$ is {\it decomposable}
with respect  to the system of states  $\{\vert b_i> \}_{i=1,2}$ ($b$-decomposable) if}
\begin{equation}
\label{E3}
v_i^b \in {\bf G}_+ \; .
\end{equation}

In such a case we can use generalization of Born's probabilistic interpretation for
a hyperbolic Hilbert space (this is a postulate!).
Numbers
$$
p_\psi^b(b_i)= \vert v_i^b \vert^2, i=1,2,
$$
are interpreted as probabilities
for values $b=b_i$ for the ${\bf G}$-quantum state $\psi.$

Thus decomposability is not a mathematical notion. This is not just linear algebraic
decomposition of a vector with respect a basis. This is a physical notion describing
the possibility of probability interpretation of a measurement over a state. As it was already
mentioned, in hyperbolic quantum mechanics a state $\psi\in {\bf E}$ is not always decomposable. Thus for an
observable $b$ there can exist a state $\psi$ such that the probabilities $p_\psi^b(b_i)$ are not well defined.
One of reasons for this can be the impossibility to perform the $b$-measurement for systems in the state $\psi.$
Such a situation is quite natural from the experimental viewpoint. Moreover, it looks surprising that in ordinary
quantum (as well as classical) theory we can measure any observable in any state. I think that this is just
a consequence of the fact that there was fixed the set of states corresponding to a rather special class of
physical observables. Thus in the hyperbolic quantum formalism for each state $\psi \in {\bf E}$
there exists its own set of observables ${\cal O}(\psi).$ And in general ${\cal O}(\psi) \not= {\cal O}(\psi).$
We cannot exclude another possibility. The set of observables ${\cal O}$ does not depend on a state $\psi.$
And the result of an individual measurement of any $b\in {\cal O}$ is well defined for any state $\psi.$ But
relative frequencies of realizations of the value $b=b_k$ do not converge to any limit. Therefore probabilities
are not well defined. Thus the principle of the statistical stabilization is violated, see [50] for details.

{\bf Remark.} Let ${\cal K}$ be a Kolmogorov probability model and let
$\psi \in S_{{\cal C}^{\rm{hyp}}}.$ Thus $\psi= \psi_C$ for some context
$C \in {\cal C}^{\rm{hyp}}.$ Let the matrix of transition probabilities ${\bf P}^{b/a}$ be double
stochastic. Then $\psi$ is decomposable with respect to both reference variables $b$ and
$a.$ Moreover, basis vectors $e_i^b= \vert b_i>$ are $a$-decomposable and vice versa.

 Suppose that a state $\psi \in {\bf E}$ is $a$-decomposable:
$$
\psi= v_1^a \vert a_1> + v_2^a \vert a_2>
$$
and the coefficients $v_i^a \in {\bf G}_+.$

We  also suppose that each state $\vert a_i>$ is decomposable with respect
to the system of states $\{\vert b_i>\}_{i=1,2}.$ We have:
\begin{equation}
\label{E4}
\vert a_1>=v_{11} \vert b_1> + v_{12} \vert b_2>,\; \;
\vert a_2>= v_{21} \vert b_1> + v_{22} \vert b_2>\;,
\end{equation}
where the coefficients $v_{ik}$ belong to ${\bf G}_+.$   We have (since both bases are orthonormal):
\begin{equation}
\label{E5}
|v_{11}|^2 + |v_{12}|^2 = 1, \; \;|v_{21}|^2 + |v_{22}|^2 = 1\;,
\end{equation}
cf. (\ref{NORM3}).
We can use the probabilistic interpretation of numbers $ p_{ik} = |v_{ik}|^2,$ namely
$p_{ik}= p_{\vert a_i>}(b_k)$ is the probability for $b=b_k$ in the state $\vert a_i>.$

Let us consider matrix $V=(v_{ik}).$
As in the complex case, the matrix $V$ is unitary, since vectors $\vert a_1>= (v_{11}, v_{12})$
and $\vert a_2>= (v_{21}, v_{22})$ are orthonormal. Hence we have normalization conditions
(\ref{E5}) and the orthogonality condition:
\begin{equation}
\label{EA}
v_{11} \bar{v}_{21} + v_{12} \bar{v}_{22}=0 \;,
\end{equation}
cf. (\ref{NORM2}).
It must be noticed that in general unitarity does not imply that $v_{ik} \in {\bf G}_+.$
The latter condition is the additional constraint on the unitary matrix $V.$
Let us consider the matrix  ${\bf P} ^{b/a} =(p_{ik}).$ This matrix is double
stochastic (since $V$ is unitary).

By using the ${\bf G}$-linear space calculation (the change of the basis) we get
$\psi= v_1^b \vert b_1> + v_2^b \vert b_2>,$
where $v_1^b = v_1^a v_{11}+ v_2^a v_{21}$ and
$v_2^b = v_1^a v_{12}+ v_2^a v_{22}.$

We remark that decomposability is not transitive. In principle $\psi$
may be not decomposable with respect to $\{\vert b_i>\}_{i=1,2},$
despite the decomposability of $\psi$ with respect to $\{\vert a_i>\}_{i=1,2}$
and the decomposability of the latter system with
respect to $\{\vert b_i>\}_{i=1,2}.$

The possibility of decomposability is based on two (totally different) conditions:
(\ref{E2}), normalization, and (\ref{E3}), positivity. Any ${\bf G}$-unitary transformation
preserves the normalization condition. Thus we get automatically that
$\vert v_1^b \vert^2 + \vert v_2^b \vert^2 =1.$ However, the condition of positivity in general is not preserved:
it can be that $v_i^b \not\in {\bf G}_+$ even if we have $v_i^a \in {\bf G}_+$ and
the matrix $V$ is ${\bf G}$-unitary.

Finally,  suppose that $\psi$ is decomposable with respect to $\{\vert b_i>\}_{i=1,2}.$
Thus  $v_k^b \in {\bf G}_+.$
Therefore coefficients $p_\psi^b(b_i) = \vert v_i^b \vert^2$ can be interpreted as
probabilities for $b=b_k$ for the ${\bf G}$-quantum state $\psi.$

Let us consider states such that coefficients $v_i^a, v_{ik}$ belong to ${\bf G}_+^*.$
We can uniquely represent them
as

$v_i^a=\pm \sqrt{p_\psi^a(a_i)} e^{j \xi_i},
v_{ik}=\pm \sqrt{p_{ik}} e^{j \gamma_{ik}}, i, k, =1,2.$

We find that
\begin{equation}
\label{E7a}
p_\psi^b(b_1) = p_\psi^a(a_1) p_{11} + p_\psi^a(a_2) p_{21} +
2 \epsilon_1 \cosh \theta_1 \sqrt{p_\psi^a(a_1) p_{11}p_\psi^a(a_2) p_{21}} \;,
\end{equation}
\begin{equation}
\label{E7b}
p_\psi^b(b_2) = p_\psi^a(a_1) p_{12} + p_\psi^a(a_2) p_{22} +
2 \epsilon_2 \cosh \theta_2 \sqrt{p_\psi^a(a_1) p_{12} p_\psi^a(a_2) p_{22}} \;,
\end{equation}
where $\theta_i = \eta+ \gamma_i$  and $\eta= \xi_1- \xi_2,
\gamma_1= \gamma_{11}- \gamma_{21}, \gamma_1= \gamma_{12}- \gamma_{22}$
and $\epsilon_i= \pm.$
To find the right relation between signs of the last terms in equations  (\ref{E7a}),
(\ref{E7b}), we use the normalization condition
\begin{equation}
\label{E7}
\vert v_2^b \vert^2 + \vert v_2^b \vert^2=1
\end{equation}
(which is a consequence of the normalization of $\psi$ and orthonormality of
the system $\{\vert b_i>\}_{i=1,2}).$
\footnote{We remark that the normalization condition (\ref{E7}) can be reduced to
relations between coefficients of the transition matrix $V.$ So it does not depend
on the original $a$-decomposition of $\psi,$ namely coefficients $v_i^a.$ Condition
of positivity, $\vert v_i^b \vert^2 \geq 0,$  could not be written
by using only coefficients of $V.$ We also need to use coefficients $v_i^a.$
Therefore it seems to be impossible to find such a class of linear transformations
$V$ that would preserve condition of positivity, ``decomposition-group" of operators.}

Equation (\ref{E7}) is equivalent to the equation:
\begin{equation}
\label{E8}
\sqrt{p_{12}p_{22}} \cosh\theta_2 \pm \sqrt{p_{11}p_{21}} \cosh\theta_2=0.
\end{equation}
Thus we have to choose opposite signs in equations (\ref{E7a}), (\ref{E7b}).
Unitarity of $V$ also implies that $\theta_1 - \theta_2 =0,$ so $\gamma_1= \gamma_2.$
We recall that in the ordinary quantum mechanics we have similar conditions, but
trigonometric functions are used instead of hyperbolic
and phases $\gamma_1$ and $\gamma_2$ are such that $\gamma_1 - \gamma_2= \pi.$

Finally, we get that unitary linear transformations in the ${\bf G}$-Hilbert space
(in the domain of  decomposable states) represent the following transformation
of probabilities:
\begin{equation}
\label{E7c}
p_\psi^b(b_1) = p_\psi^a(a_1) p_{11} + p_\psi^a(a_2) p_{21} \pm
2 \epsilon_1 \cosh \theta_1 \sqrt{p_\psi^a(a_1) p_{11}p_\psi^a(a_2) p_{21}} \;,
\end{equation}
\begin{equation}
\label{E7d}
p_\psi^b(b_2)= p_\psi^a(a_1) p_{12} + p_\psi^a(a_2) p_{22} \mp
2 \epsilon_2 \cosh \theta_2 \sqrt{p_\psi^a(a_1) p_{12} p_\psi^a(a_2) p_{22}}  .
\end{equation}
This is a kind of hyperbolic interference.

\section{Complex amplitudes of probabilities in the case of multivaried reference variables}

The general case of random variables taking $n\geq 2$ different values can be (inductively) reduced
to the case of dichotomous random variables (cf., e.g., Mackey [14] who also reduced the study
of arbitrary observables to the study of dichotomous variables - questions). We consider two incompatible
random variables taking $n$ values: $b=b_1, \ldots, b_n$ and
$a=a_1, \ldots, a_n.$

{\bf Lemma 7.} {\it{Let $B, C, D_1, D_2 \in {\cal F}, {\bf P}(C)\ne 0$ and $D_1\cap D_2=\emptyset.$ Then}}
\begin{equation}
\label{F1}
{\bf P}(B(D_1 \cup D_2)/C)={\bf P}(BD_1/C)+{\bf P}(BD_2/C)
\end{equation}

{\bf Proposition 5.} (The formula of total probability) {\it{Let conditions of Lemma 7 hold
and let ${\bf P}(D_jC)\ne 0.$ Then}}

\begin{equation}
\label{F2}
{\bf P}(B(D_1 \cup D_2)/C)={\bf P}(B/D_1C){\bf P}(D_1/C)+{\bf P}(B/D_2C){\bf P}(D_2/C)
\end{equation}

{\bf Proposition 6.} (Contextual formula of total probability) {\it{Let conditions of Proposition 5 hold true
and let ${\bf P}(BD_j)\ne 0, j=1,2.$ Then
\begin{equation}
\label{F3}
{\bf P}(B(D_1 \cup D_2)/C)={\bf P}(B/D_1){\bf P}(D_1/C)+{\bf P}(B/D_2){\bf P}(D_2/C)+
\end{equation}
\[2\lambda(B/\{D_1, D_2\}, C) \sqrt{{\bf P}(B/D_1){\bf P}(D_1/C){\bf P}(B/D_2){\bf P}(D_2/C)},\]
where
\begin{equation}
\label{LA}
\lambda(B/\{D_1, D_2\}, C)=\frac{\delta(B/\{D_1, D_2\}, C)}{2\sqrt{{\bf P}(B/D_1){\bf P}(D_1/C)
{\bf P}(B/D_2){\bf P}(D_2/C)}}
\end{equation}
and }
\[\delta(B/\{D_1, D_2\}, C)={\bf P}(B(D_1 \cup D_2)/C) - \sum_{j=1}^{2} {\bf P}(B/D_j){\bf P}(D_j/C)\]
\[=\sum_{j=1}^{2} {\bf P}(D_j/C) ({\bf P}(B/D_j C)-{\bf P}(B/D_j))\]}

In the construction of a Hilbert space representation of contexts for multivalued observables there
will be used the following combination of formulas (\ref{F1}) and (\ref{F3}).

{\bf Lemma 8.} {\it Let conditions of Lemma 7 hold and let ${\bf P}(BD_1), {\bf P}(CD_1)$ and ${\bf P}(BD_2 C)$ be strictly positive. Then
\begin{equation}
\label{F5}
{\bf P}(B(D_1 \cup D_2)/C)={\bf P}(B/D_1){\bf P}(D_1/C)+ {\bf P}(BD_2/C)
\end{equation}
\[+ 2 \mu(B/\{D_1, D_2\}, C) \sqrt{{\bf P}(B/D_1){\bf P}(D_1/C){\bf P}(BD_2/C)}\]
where} $\mu(B/\{D_1, D_2\}, C)=\frac{{\bf P}(B(D_1 \cup D_2)/C)-{\bf P}(B/D_1){\bf P}(D_1/C)-{\bf P}(BD_2/C)}{2\sqrt{{\bf P}(B/D_1){\bf P}(D_1/C){\bf P}(BD_2/C)}
}$

\medskip

Suppose that coefficients of statistical disturbance $\mu$ and $\lambda$ are bounded
by 1. Then we can represent them in the trigonometric form:
\[
\lambda(B/\{D_1, D_2\}, C)=\cos \theta (B/\{D_1, D_2\}, C)
\]
\[
\mu(B/\{D_1, D_2\}, C) =\cos \gamma (B/\{D_1, D_2\}, C)
\]

By inserting these $cos$-expressions in (\ref{F3}) and (\ref{F5}) we obtain trigonometric transformations of
probabilities. We have (by Lemma 8):
$$
{\bf P}(B_x/C)={\bf P}(B_x(A_1 \cup \ldots \cup A_n)/C)
$$
$$
={\bf P}(B_x/A_1){\bf P}(A_1/C)
+ {\bf P}(B_x(A_2 \cup \ldots \cup A_n)/C)
$$
$$
+ 2\mu(B_x/\{A_1, A_2 \cup \ldots \cup A_n\}, C) \sqrt{ {\bf P}(B_x/A_1){\bf P}(A_1/C){\bf P}(B_x(A_2 \cup \ldots \cup A_n)/C)},
$$
where
$$
\mu(B_x/\{A_1, A_2 \cup \ldots \cup A_n\}, C)
$$
$$
=\frac{{\bf P}(B_x(A_1 \cup \ldots \cup A_n)/C) - {\bf P}(B_x/A_1){\bf P}(A_1/C)-{\bf P}(B_x(A_2 \cup \ldots \cup A_n)/C)}
{2\sqrt{{\bf P}(B_x/A_1){\bf P}(A_1/C){\bf P}(B_x(A_2 \cup \ldots \cup A_n)/C))}} .
$$
Suppose that the coefficients of statistical disturbance are relatively small for all $x \in X:$
$|\mu(B_x/\{A_1, A_2 \cup \ldots \cup A_n\}, C)|\leq 1.$
Then we can represent these coefficients as
$$
\mu(B_x/\{A_1, A_2 \cup \ldots \cup A_n\}, C) = \cos \gamma (B_x/\{A_1, A_2 \cup \ldots \cup A_n\}, C).
$$
Thus the probability ${\bf P}(B_x/C)\equiv {\bf P}(B_x(A_1 \cup \ldots \cup A_n)/C)$
can be represented as the square of the absolute value of the complex amplitude:
$$
\psi_C(x)\equiv \psi_C^{(1)}(x)=\sqrt{{\bf P}(B_x/A_1){\bf P}(A_1/C)} +
e^{i\gamma_C^{(1)}(x)} \sqrt{{\bf P}(B_x(A_2 \cup \ldots \cup A_n)/C)},
$$
where the phase $\gamma_C^{(1)} (x) \equiv \gamma(B_x/\{A_1, A_2 \cup \ldots \cup A_n\}, C).$
In the same way the probability in the second summand can be represented as:
$$
{\bf P}(B_x(A_2 \cup \ldots \cup A_n)/C)=
{\bf P}(B_x/A_2){\bf P}(A_2/C)+ {\bf P}(B_x(A_3 \cup \ldots \cup A_n)/C) +
$$
$$
2\mu(B_x/\{A_2, A_3 \cup \ldots \cup A_n\}, C)
\sqrt{{\bf P}(B_x/A_2){\bf P}(A_2/C) {\bf P}(B_x(A_3 \cup \ldots \cup A_n)/C)},
$$
where
$$
\mu(B_x/\{A_2, A_3 \cup \ldots \cup A_n\}, C)
$$
$$
=\frac{{\bf P}(B_x(A_2 \cup \ldots \cup A_n)/C)-{\bf P}(B_x/A_2){\bf P}(A_2/C)-{\bf P}(B_x(A_3 \cup \ldots \cup A_n)/C)}{2\sqrt{{\bf P}(B_x/A_2){\bf P}(A_2/C){\bf P}(B_x(A_3 \cup \ldots \cup A_n)/C)}}.
$$

By supposing that these coefficients of statistical disturbance are bounded by 1 we represent the probability as the square of the absolute value of the complex amplitude:
$$
\psi_C^{(2)}(x)=\sqrt{{\bf P}(B_x/A_2){\bf P}(A_2/C)} + e^{i \gamma_C^{(2)}(x)} \sqrt{{\bf P}(B_x(A_3 \cup \ldots \cup A_n)/C)},
$$
where $\gamma_C^{(2)}(x)=\pm \arccos \mu (B_x/\{A_2, A_3, \cup \ldots \cup A_n\}, C).$ On the $jth$ step we represent ${\bf P}(B_x(A_j \cup \ldots \cup A_n)/C)$ as the square of the absolute value of the complex amplitude
\[\psi_C^{(j)}(x)=\sqrt{{\bf P}(B_x/A_j){\bf P}(A_j/C)} + e^{i \gamma_C^{(j)}(x)} {\sqrt{{\bf P}(B_x(A_{j+1}\cup \ldots \cup A_n)/C)}},\]
where $\gamma_C^{(j)}(x)$ is the phase of the coefficient
$$
\mu (B_x/\{A_j, A_{j+1} \cup \ldots \cup A_n\}, C)
$$
$$
=\frac{{\bf P}(B_x(A_j \cup \ldots \cup A_n)/C)-{\bf P}(B_x/A_j) {\bf P}(A_j/C)-{\bf P}(B_x (A_{j+1}\cup \ldots \cup A_n)/C)}{2\sqrt{{\bf P}(B_x/A_j) {\bf P}(A_j/C) {\bf P}(B_x (A_{j+1} \cup \ldots \cup A_n)/C)}}.$$

It is supposed that at each step we obtain coefficients $|\mu|$ bounded by 1.
At the step $j=n-1$ we should represent the probability ${\bf P}(B_x(A_{n-1} \cup A_n)/C).$
Here we can already totally eliminate the $C$-contextuality for $B_x:$
\[{\bf P}(B_x(A_{n-1} \cup A_n)/C)={\bf P}(B_x/A_{n-1}) {\bf P} (A_{n-1}/C)+
{\bf P}(B_x/A_{n}) {\bf P} (A_{n}/C) \]
\[+2\lambda (B_x/\{ A_{n-1}, A_n\})
\sqrt{{\bf P}(B_x/A_{n-1}) {\bf P}(A_{n-1}/C) {\bf P}(B_x/A_n) {\bf P}(A_n/C)},\]
where the coefficient of statistical disturbance $\lambda$ was defined by (\ref{LA}).
And if $|\lambda|$ is bounded by 1 then we can represent the probability as the square
of the absolute value of the complex amplitude:
$$
\psi_C^{(n-1)}(x)=\sqrt{{\bf P}(B_x/A_{n-1}) {\bf P}(A_{n-1}/C)}  +
e^{i \theta_C(x)} \sqrt{{\bf P}(B_x/A_n) {\bf P}(A_n/C)},
$$
where $\theta_C(x)=\pm \arccos \lambda (x/\{A_{n-1}, A_n\}, C).$

We have:
$$
\psi_C^{(j)}(x)=\sqrt{{\bf P}(B_x(A_j \cup \ldots \cup A_n)/C)} \; e^{i \alpha_C^{(j)}(x)},
$$
where $\alpha_C^{(j)}(x)=\arg \psi_C^{(j)}(x)
= \arccos \frac{M_j}{N_j},$
where $ M_j=\sqrt{{\bf P}(B_x/A_j) {\bf P}(A_j/C)}$\\
$+\mu (B_x/\{A_j, A_{j+1} \cup \ldots \cup A_n\}, C)\sqrt{{\bf P}(B_x(A_{j+1}\cup \ldots \cup A_n)/C)},$\\
$
N_j=\sqrt{{\bf P}(B_x (A_j \cup \ldots \cup A_n) /C)}.
$
Finally, we have:
$$
\alpha_C^{(n-1)}(x)=\arg \psi_C^{(n-1)}(x)
$$
$$
=\arccos \frac{\sqrt{{\bf P}(B_x/A_{n-1}) {\bf P}(A_{n-1}/C)} +
\lambda (B_x/\{A_{n-1}, A_n\},C) \sqrt{{\bf P}(B_x/A_n) {\bf P}(A_n/C)}}
{\sqrt{{\bf P}(B_x(A_{n-1} \cup A_n)/C)}}.
$$
Thus we have:
$$
\psi_C(x)=\sqrt{{\bf P}(B_x/A_1) {\bf P}(A_1/C)} + e^{i [\gamma_C^{(1)}(x) - \alpha_C^{(2)}(x)]}
\psi_C^{(2)}(x)
$$
$$
=\sqrt{{\bf P}(B_x/A_1) {\bf P}(A_1/C)} + e^{i \beta_C^{(2)}(x)} \sqrt{{\bf P}(B_x/A_2) {\bf P}(A_2/C)}
$$
$$
+ e^{i\beta_C^{(3)}(x)} \psi_C^{(3)}(x),
$$
where
$$
\beta_C^{(2)}(x)=\gamma_C^{(1)}(x) - \alpha_C^{(2)}(x), \beta_C^{(3)}(x)=\beta_C^2(x) +
\gamma_C^{(2)}(x) - \alpha_C^{(3)}(x).
$$
Finally, we obtain:
$$
\psi_C(x)=\sum_{j=1}^n e^{i\beta_C^{(j)}(x)} \sqrt{{\bf P}(B_x/A_j) {\bf P}(A_j/C)}
$$
with $\beta_C^{(1)}(x)=0$ (this is just due to our special choice of a representation)
and $\beta_C^{(n)}(x)=\beta_C^{(n-1)}(x) + \theta_C (x).$

Thus by inductive splitting of multivalued variables into dichotomous
variables we represented contextual probabilities by complex amplitudes
$\psi_C(x).$

By using the standard in this paper
symbols
$p(x/y)={\bf P}(B_x/A_y)$ and $p_C^b(x)={\bf P}(B_x/C), p_C^a(y)={\bf P}(A_y/C)$
we write
$$
\psi_C(x)=\sum_y e^{i \beta_C^{(y)}(x)} \sqrt{p_C^a(y) p(x/y)}.
$$
In particular, for $n=3$ we have
$$
\psi_C(x)=\sqrt{p_C^a(a_1) p(x/a_1)} + e^{i \beta_C^{(2)} (x)} \sqrt{p_C^a(a_2) p(x/a_2)},
+ e^{i \beta_C^{(3)}(x)} \sqrt{p_C^a(a_3) p(x/a_3)} ,
$$
where
$$
\beta_C^{(2)}(x)=\gamma_C^{(1)}(x) - \alpha_C^{(2)}(x),
\beta_C^{(3)}(x)=\beta_C^{(2)}(x)+ \theta_C(x).
$$
We remark that each phase
$\beta_C^{(j)}(x)$ depends on all three $a$-contexts, $A_1, A_2, A_3.$ So
we cannot use the symbol $\beta_C(x/y).$ In $\beta_C^{(y)}(x)$ the $y$ is just the summation index;
in fact, $\beta_C^{(y)}(x)\equiv \beta_C^{(y)}(x/A_1, A_2, A_3).$ We remark that the probability
$p_C^b(x)$ can be represented as
$$
p_C^b(x)=|\psi_C(x)|^2=\sum_y p_C^a(y) p(x/y)
$$
$$
+ 2\sum_{y_1<y_2}\cos [\beta_C^{(y_2)}(x)-\beta_C^{(y_1)}(x)] \sqrt{p_C^a (y_1) p_C^a (y_2) p(x/y_1) p(x/y_2)} .
$$

We can proceed in the same way as in the case of dichotomous random variables.

\bigskip

I would like to thank L. Ballentine, S. Gudder, A. Holevo, P. Lahti, B Hiley,  C. Fuchs,
A. Peres, I. Volovich, R. Gill, K. Hess, W. Philipp, L. Accardi, A. Aspect, G. `t Hooft,
J. Bub, T. Maudlin,  H. Rauch, G. Emch, V. Belavkin, A. Leggett, I. Helland, P. Kwait, G. Adenier,
for discussions on probabilistic foundations of quantum theory. This paper was partially supported by EU-Network
"QP and Applications'' and Nat. Sc. Found., grant N PHY99-07949 at KITP, Santa-Barbara,
visiting professor fellowship
at Russian State Humanitarian University and the Profile Mathematical
Modeling of V\"axj\"o University.

\section{Appendix on incompatible random variables}

 {\bf Proposition 7.}
 {\it Let $\{A_j\}$ and $\{ B_k\}$ be two families of subsets of some set $\Omega$ and
 $\Omega= \cup_j A_j = \cup_k B_k$ and
 let
 \begin{equation}
 \label{I}
 A_j B_k\ne \emptyset
 \end{equation}
 for any pair $(j,k).$  Then
 \begin{equation}
  \label{I1}
{\mbox{Neither}}\;A_j \subset B_k \;{\mbox {nor}} \; \; B_k \subset A_j
  \end{equation}
  for any pair $(j,k).$ If $n=2$ then conditions (\ref{I}) and (\ref{I1}) are equivalent.}

  {\bf{Proof.}} Let (\ref{I}) hold true. Suppose that there exists $(j, k)$ such that $A_j \subset B_k.$
Thus we should have $A_j B_i= \emptyset$ for any $i\ne k.$ Let (\ref{I1}) hold true and
let $n=2:{\cal A}=\{A_1, A_2=\Omega\setminus A_1\}$ and ${\cal B}=\{B_1, B_2=\Omega\setminus B_1\}.$
Suppose that, e.g., $A_1 B_1=\emptyset.$ Then we should have $A_1 \subset B_2.$

If $n\neq 3$ then in general the condition (\ref{I1}) does not imply the condition (\ref{I}).
We can consider the following example. Let $\Omega=\{\omega_1, \ldots, \omega_7\}$
and let $A_1=\{\omega_1, \omega_2 \omega_3\}, A_2=\{\omega_4, \omega_5\},
A_3=\{\omega_6, \omega_7\}$ and $B_1=\{\omega_1, \omega_4\},
B_2=\{\omega_2, \omega_5, \omega_6\}, B_3=\{\omega_3, \omega_7\}.$
Here (\ref{I1}) holds true but $A_2 B_3\not= \emptyset.$

\medskip

\centerline{\bf REFERENCES}

\medskip

1. A. N. Kolmogoroff, {\it Grundbegriffe der Wahrscheinlichkeitsrechnung,}
Springer Verlag, Berlin, 1933; reprinted:
{\it Foundations of the Probability Theory,}
Chelsea Publ. Comp., New York, 1956.

2.  D. Hilbert, J. von Neumann, L. Nordheim, {\it Math. Ann.}, {\bf 98}, 1-30 (1927).

3. P. A. M.  Dirac, {\it The Principles of Quantum Mechanics,}
Oxford Univ. Press, 1930.

4. W. Heisenberg, {\it Physical principles of quantum theory,}
Chicago Univ. Press, 1930.

5. J. von Neumann, {\it Mathematical foundations
of quantum mechanics,} Princeton Univ. Press, Princeton, N.J., 1955.

6. A. Einstein, B. Podolsky, N. Rosen, {\it Phys. Rev.} {\bf 47}, 777--780 (1935).

7. N. Bohr, {\it Phys. Rev.} {\bf 48}, 696-702 (1935).

8. D. Bohm, {\it Quantum theory,} Prentice-Hall,
Englewood Cliffs, New-Jersey, 1951.

9. A. Lande, {\it Foundations of quantum theory,} Yale Univ. Press, 1955.

10. A. Lande, {\it New foundations of quantum mechanics,} Cambridge Univ. Press, Cambridge, 1965.

11. A. S. Wightman, Hilbert's sixth problem: mathematical treatment of the axioms of physics,
{\it Proc. Symposia in Pure Math.,} {\bf 28}, 147-233 (1976).

12. L. De Broglie, {\it The current interpretation of wave mechanics,
critical study.} Elsevier Publ., Amsterdam-London-New York, 1964.

13. J. S. Bell, {\it Speakable and unspeakable in quantum mechanics,} Cambridge Univ. Press, 1987.

14. G. W. Mackey, {\it Mathematical foundations of quantum mechanics,}
W. A. Benjamin INc, New York, 1963.

15.  S. Kochen and E. Specker, {\it J. Math. Mech.}, {\bf 17}, 59-87 (1967).

16. L. E. Ballentine, {\it Rev. Mod. Phys.}, {\bf 42}, 358--381 (1970).

17.  G. Ludwig, {\it Foundations of quantum mechanics,} Springer,
Berlin, 1983.

18. E. B. Davies, J. T. Lewis, {\it Comm. Math. Phys.} {\bf 17}, 239-260 (1970).

19. E. Nelson, {\it Quantum fluctuation,} Princeton Univ. Press, Princeton, 1985.

20. D.  Bohm  and B. Hiley, {\it The undivided universe:
an ontological interpretation of quantum mechanics,} Routledge and Kegan Paul, London, 1993.

21. S. P. Gudder,
{\it Trans. AMS} {\bf 119}, 428-442 (1965).

22. S. P. Gudder, {\it Axiomatic quantum mechanics and generalized probability theory,}
Academic Press, New York, 1970.

23. S. P. Gudder, {\it ``An approach to quantum probability''} in
{\it Foundations of Probability and Physics,} edited by  A. Yu. Khrennikov,
Quantum Prob. White Noise Anal., 13,  WSP, Singapore, 2001, pp. 147-160.

24. R. Feynman and A. Hibbs, {\it Quantum Mechanics and Path Integrals,}
McGraw-Hill, New-York, 1965.

25. J. M. Jauch, {\it Foundations of Quantum Mechanics,} Addison-Wesley,
Reading, Mass., 1968.

26. A. Peres, {\em Quantum Theory: Concepts and Methods,} Dordrecht, Kluwer Academic, 1994.

27. L. Accardi, {\it ``The probabilistic roots of the quantum mechanical paradoxes''} in
{\em The wave--particle dualism.  A tribute to Louis de Broglie on his 90th
Birthday,} edited by  S. Diner, D. Fargue, G. Lochak and F. Selleri,
D. Reidel Publ. Company, Dordrecht, 1984, pp. 297--330.

28. L. Accardi, {\it Urne e Camaleoni: Dialogo sulla realta,
le leggi del caso e la teoria quantistica,} Il Saggiatore, Rome, 1997.

29.  L. E. Ballentine, {\it Quantum mechanics,} Englewood Cliffs,
New Jersey, 1989.

30. L. E. Ballentine,  {\it ``Interpretations of probability and quantum theory'',} in
{\it Foundations of Probability and Physics,} edited by  A. Yu. Khrennikov,
 Q. Prob. White Noise Anal.,  13,  WSP, Singapore, 2001, pp. 71-84.

31. A. S. Holevo, {\it Probabilistic and
statistical aspects of quantum theory,} North-Holland, Amsterdam,  1982.

32. A. S. Holevo, {\it Statistical structure of quantum theory,} Springer,
Berlin-Heidelberg, 2001.

33. P. Busch, M. Grabowski, P. Lahti, {\it Operational Quantum Physics,}
Springer Verlag,Berlin, 1995.

34. A. Yu. Khrennikov (editor), {\it Foundations of Probability and Physics,}
Q. Prob. White Noise Anal.,  13,  WSP, Singapore, 2001.

35. A. Yu. Khrennikov (editor), {\it Quantum Theory: Reconsideration
of Foundations,} Ser. Math. Modeling, 2, V\"axj\"o Univ. Press,  2002.

36. A. Yu. Khrennikov (editor), {\it Foundations of Probability and Physics}-2,
Ser. Math. Modeling, 5, V\"axj\"o Univ. Press,  2003.

37. A. Yu. Khrennikov (editor),  {\it Quantum Theory: Reconsideration
of Foundations}-2,  Ser. Math. Modeling, 10, V\"axj\"o Univ. Press,  2004.

38. A. Yu. Khrennikov, {\it J. Phys.A: Math. Gen.} {\bf 34}, 9965-9981 (2001).

39.  A . Yu. Khrennikov, {\it Il Nuovo Cimento} {\bf B 117},  267-281 (2002).

40.  A. Yu. Khrennikov, {\it J. Math. Phys.} {\bf 43}, 789-802 (2002).

41.   A. Yu. Khrennikov, {\it Information dynamics in cognitive, psychological and
anomalous phenomena,} Ser. Fundamental Theories of Physics, Kluwer, Dordreht, 2004.

42. A. Yu. Khrennikov,  {\it J. Math. Phys.} {\bf 44},  2471- 2478 (2003).

43. A. Yu. Khrennikov, {\it Phys. Lett. A} {\bf 316}, 279-296 (2003).

44. A. Yu. Khrennikov, {\it Advances in Applied Clifford Algebras}
{\bf 13}(1),  1-9 (2003).

45.  A. Yu. Khrennikov,
{\it Annalen  der Physik} {\bf 12},  575-585 (2003).

46.  A. N. Shiryayev, {\it Probability,} Springer, New York-Berlin-Heidelberg, 1991.

47. E. Conte, O. Todarello,  A. Federici, F. Vitiello, M. Lopane,  A. Yu. Khrennikov,
{\it ``A preliminary evidence of quantum-like behaviour in measurements of mental states''} in
{\it Quantum Theory: Reconsideration
of Foundations,} edited by A. Yu. Khrennikov,  Ser. Math. Modeling, 10,
 V\"axj\"o Univ. Press, 2004, pp. 679-702.

48. H. Atmanspacher, H. Primas, {\it ``Epistemic and ontic quantum realities''}, in
{\it Foundations of Probability and Physics}-3, edited by A. Yu. Khrennikov, AIP Conference Proceedings, 2005.

49.   A. Yu. Khrennikov, {\it Supernalysis, } Nauka, Fizmatlit, Moscow, 1997 (in
Russian). English translation: Kluwer, Dordreht, 1999.

50. A. Yu. Khrennikov, {\it Interpretations of Probability,}
VSP Int. Sc. Publishers, Utrecht/Tokyo, 1999 (second edition, 2004).

\end{document}